\documentclass[10pt,english,notitlepage,a4paper,superscriptaddress,floatfix,longbibliography,nofootinbib,pre]{revtex4-1}

\usepackage[T1]{fontenc}
\usepackage{geometry}
\usepackage{float}
\usepackage{amsmath}
\usepackage{amssymb}
\usepackage{graphicx}
\usepackage{esint}
\usepackage{xcolor}
\usepackage{comment}
\usepackage[unicode=true,pdfusetitle,
 bookmarks=true,bookmarksnumbered=false,bookmarksopen=false,
 breaklinks=false,pdfborder={0 0 1},backref=false,colorlinks=false]
 {hyperref}

\makeatletter

\floatstyle{ruled}
\newfloat{algorithm}{H}{loa}
\providecommand{\algorithmname}{Algorithm}
\floatname{algorithm}{\protect\algorithmname}

\pdfoutput=1
\usepackage{makeidx}\usepackage{amsfonts}

\usepackage{xcolor}

\newcommand{\Causalityname}[1]{Causal Variational Approach}
\newcommand{\Causalityacronym}[1]{CVA}
\newcommand{\assignmentop}{\leftarrow}

\makeatother

\begin{document}
\title{Inference in conditioned dynamics through causality restoration}
\author{Alfredo Braunstein}
\affiliation{DISAT, Politecnico di Torino, Corso Duca Degli Abruzzi 24, 10129 Torino}
\affiliation{Collegio Carlo Alberto, P.za Arbarello 8, 10122, Torino, Italy}
\affiliation{Italian Institute for Genomic Medicine, IRCCS Candiolo, SP-142, I-10060 Candiolo (TO), Italy}
\author{Giovanni Catania}
\affiliation{Departamento de F\'isica T\'eorica I, Universidad Complutense, 28040 Madrid, Spain}
\author{Luca Dall'Asta}
\affiliation{DISAT, Politecnico di Torino, Corso Duca Degli Abruzzi 24, 10129 Torino}
\affiliation{Collegio Carlo Alberto, P.za Arbarello 8, 10122, Torino, Italy}
\affiliation{Italian Institute for Genomic Medicine, IRCCS Candiolo, SP-142, I-10060 Candiolo (TO), Italy}
\author{Matteo Mariani}
\email{matteo.mariani@polito.it}
\affiliation{DISAT, Politecnico di Torino, Corso Duca Degli Abruzzi 24, 10129 Torino}
\author{Anna Paola Muntoni}
\affiliation{DISAT, Politecnico di Torino, Corso Duca Degli Abruzzi 24, 10129 Torino}
\affiliation{Italian Institute for Genomic Medicine, IRCCS Candiolo, SP-142, I-10060 Candiolo (TO), Italy}

\begin{abstract}
Estimating observables from conditioned dynamics is typically computationally hard. While obtaining independent samples efficiently from unconditioned dynamics is usually feasible, most of them do not satisfy the imposed conditions and must be discarded.
On the other hand, conditioning breaks the causal properties of the dynamics, which ultimately renders the sampling of the conditioned dynamics non-trivial and inefficient. 
In this work, a Causal Variational Approach is proposed, as an approximate method to generate independent samples from a conditioned distribution. The procedure relies on learning the parameters of a generalized dynamical model that optimally describes the conditioned distribution in a variational sense. The outcome is an effective and unconditioned dynamical model from which one can trivially obtain independent samples, effectively restoring the causality of the conditioned dynamics. The consequences are twofold: on the one hand, it allows one to efficiently compute observables from the conditioned dynamics by averaging over independent samples; on the other hand, the method provides an effective unconditioned distribution that is easy to interpret. This approximation can be applied virtually to any dynamics. The application of the method to epidemic inference is discussed in detail. The results of direct comparison with state-of-the-art inference methods, including the soft-margin approach and mean-field methods, are promising. 
\end{abstract}
\maketitle

\section{Introduction}

The method we will present is rather general and applies to a wide family of stochastic processes. We will thus first describe it below in complete generality, and delay its description for a specific important application (namely the risk assessment problem in epidemic spreading processes) to the following section.  
Let us denote by $\mathbb{P}\left[\mathbf{x}\right]=\mathbb{P}\left[{x}\left(0\right),...,{x}\left(k\Delta t\right)\right]$ 
the probability
distribution of trajectories $\mathbf{x}$ of a (known) dynamical model. 
Given a (hidden) realization $\mathbf{x}^{*}$, consider a set of observations
$\mathcal{O}=({O}_1,\dots,{O}_M)$ sampled from a (known) conditional distribution
$\mathbb{\ensuremath{P}}\left[\mathcal{O}|\mathbf{x}^{*}\right]$.
The scope of this work is to devise an efficient method to infer information about $\mathbf{x}^{*}$ given $\mathcal{O}$, in particular, to be able to estimate averages over the posterior distribution
\begin{equation}
\mathbb{P}\left[\mathbf{x}|\mathcal{O}\right]=\mathbb{P}\left[\mathbf{x}\right]\mathbb{P}\left[\mathcal{O}|\mathbf{x}\right]\mathbb{P}\left[\mathcal{O}\right]^{-1}.\label{eq:post}
\end{equation}
Although it might be generally feasible to sample efficiently from the prior
$\mathbb{P}\left[\mathbf{x}\right]$, sampling from $\mathbb{P}\left[\mathbf{x}|\mathcal{O}\right]$
is normally difficult. A naive approach is given by importance sampling \cite{newman_monte_1999,mackay2003information},
that consists in evaluating the average of a function $f$ by first
generating $M$ independent samples $\mathbf{x}^{1},\dots,\mathbf{x}^{M}$
from $\mathbb{P}\left[\mathbf{x}\right]$ and then computing 
\begin{equation}
\left\langle f\right\rangle \approx\frac{\sum_{\mu=1}^{M}f\left(\mathbf{x}^{\mu}\right) \mathbb{P}\left[\mathcal{O}|\mathbf{x}^{\mu}\right]}{\sum_{\mu=1}^{M} \mathbb{P}\left[\mathcal{O}|\mathbf{x}^{\mu}\right]}.
\end{equation} 
Unfortunately, this method is impractical when observations deviate
significantly from the typical case, as for the case in which $\mathbb{P}\left[\mathcal{O}|\mathbf{x}^{\mu}\right]$
becomes very small (or even zero), rendering the convergence of $\left\langle f\right\rangle$ to the true average value inefficient. 

One reason for which sampling from $\mathbb{P}\left[\mathbf{x}\right]$
is usually feasible is that the causal structure induced by the dynamical
nature of the stochastic process can be exploited to efficiently generate
trajectories. The \emph{causal} property of the stochastic dynamics
lies in the fact that the state of the system at a given time depends naturally (in a stochastic way) on states at previous times. When considering
discrete time-steps or epochs $0,\Delta t,2\Delta t,\dots$ (in the following discussion, for simplicity of notation, we will assume $\Delta t=1$), this property
implies that the distribution of trajectories of the stochastic dynamics
assumes the following factorized form:
\begin{equation}
\mathbb{P}\left[{x}(0),\dots,{x}(T)\right] = \,\prod_{t=0}^{T}\mathbb{P}\left[{x}(t)|{x}(t-1),\dots,{x}(0)\right],\label{eq:causal}
\end{equation} 
where in the $t=0$ term the conditioning part is empty and thus the probability is unconditioned. In most models, it is computationally simple (or at least feasible) to sample $x(t)$ from $\mathbb{P}\left[x(t)|x(t-1),\dots,x(0)\right]$, implying that \eqref{eq:causal} can be exploited to generate trajectories by sequentially sampling $x(0)$, then $x(1)$, etc.

The intrinsic difficulty associated with sampling from the conditioned distribution $\mathbb{P}\left[\mathbf{x}|\mathcal{O}\right]$
is a consequence of \emph{causality breaking} \cite{biroli_kurchan} induced by the addition of the extra information in $\mathcal{O}$. 
In general, $\mathbb{P}\left[x(t)|x(t-1),\dots,x(0)\right]$ and $\mathbb{P}\left[x(t)|x(t-1),\dots,x(0),\mathcal{O}\right]$ are very different objects. For example, even if the former is time and space invariant, the latter is generally not, because this symmetry is typically broken by the observations. This difference ultimately implies that we cannot sample from the posterior distribution sequentially as in the unconditioned case \eqref{eq:causal}.
Although we can write an exact expression similar to \eqref{eq:causal},
\begin{equation}
\mathbb{P}\left[\mathbf{x}|\mathcal{O}\right]=\prod_{t=0}^T \mathbb{P}\left[{x}(t)|{x}(t-1),\dots,{x}(0),\mathcal{O}\right]\label{eq:causal_post},
\end{equation} sampling from \eqref{eq:causal_post} is unfortunately still problematic. 
Indeed, the expression for $\mathbb{P}\left[x(t)|x(t-1),\dots,x(0),\mathcal{O}\right]$ is in general extremely difficult to compute and involves a marginalization over times $t'>t$ (with an exponential number of terms):
\begin{align}
\mathbb{P}\left[x(t)|x(t-1),\dots,x(0),\mathcal{O}\right] &\propto\sum_{x(t+1),\dots,x(T)} \prod_{t'=0}^{T}\mathbb{P}\left[x(t')|x(t'-1),\dots,x(0)\right]\mathbb{P}\left[\mathcal{O}|x(T),\dots,x(0)\right]. 
\end{align}
This dependence on future times is in our opinion the real source of the causality breaking phenomenon. 

When dynamics are unconditioned, i.e. causality applies, information is intuitively flowing from past to future.  Although it is a very intuitive concept, the study of information flow is actually rather involved and it opens to interesting insights into the collective interactions among agents in agent-based systems. We refer the interested reader to\cite{james2018modes,sattari2022}.
The approach proposed here, called \emph{\Causalityname{}} (\Causalityacronym{}), aims at
providing a variational approximation of the posterior distribution
$\mathbb{P}\left[\mathbf{x}|\mathcal{O}\right]$, for which 
causality features are restored and, therefore, independent samples can be efficiently generated from it. 
In particular, we propose to approximate $\mathbb{P}\left[\mathbf{x}|\mathcal{O}\right]\approx Q(\mathbf{x})$, where
$Q\left(x(0),\dots,x(T)\right)=\prod_{t=0}^Tq_t\left(x(t)|x(t-1),\dots x(0)\right)$. 
This approach is formally exact. Indeed, if we set each $q_t\left(x(t)|x(t-1),\dots x(0)\right)=\mathbb{P}\left[x(t)|x(t-1),\dots,x(0),\mathcal{O}\right]$, 
we would recover the exact posterior, due to equation \eqref{eq:causal_post}. However, this is in practice unfeasible, because it would require $q_t$ to depend on a huge (i.e. exponential in the size of the system) number of parameters. The general idea of CVA method is to restrict the functional space of $Q$ assuming the $q_t(x\left(t \right)| x \left( t-1 \right),\dots , x \left( 0 \right))$
to have the same broad functional form of the unconstrained prior distribution $\mathbb{P}\left[{x}(t)|{x}(t-1),\dots,{x}(0)\right]$, retaining then the ability to efficiently compute it and sample from it, but generalizing it by the addition of extra parameters. This generalization will naturally allow for the spatial and/or time heterogeneity that is present in the corresponding terms in the posterior, and will be explained in detail for the specific models in the next sections. In particular, we chose in the approximating distribution of CVA to maintain the following properties (if they are present) of the prior distribution:
\begin{enumerate}
    \item \textit{Spatial Independence}. In agent-based models \cite{macal_agent-based_2009} on $N$ agents, $x(t)=\left(x_1(t),\dots,x_N(t)\right)$ and most dynamical processes satisfy a spatial conditional independence property \cite{dawid_conditional_independence_1979}, namely that: 
    \begin{equation}
    \mathbb{P}[x(t)|x(t-1),\dots,x(0)] =\prod_{i=1}^{N}\mathbb{P}[x_{i}(t)|x(t-1),...,x(0)],\label{eq:causal2}
    \end{equation} 
    As this property is often crucial for efficient sampling from $\mathbb{P}\left[\mathbf{x}\right]$, CVA maintains it on the approximating $q_t$ i.e. $q_t({x}(t)|{x}(t-1),\dots,{x}(0))= \prod_i q^i_t(x^i(t)|{x}(t-1),\dots,{x}(0))$. 
    \item \textit{Local interactions}. Each variable of the prior process, moreover, might depend only on a restricted (local) set of variables on a given contact network; we choose to preserve this dependence in $q_t$. Note that this property is in general not present in the posterior. 

   \item \textit{Markovianity}. If the prior distribution defines a memory-less stochastic process\cite{markov_chains}, namely
    \begin{equation}
    \mathbb{P}\left[x(t)|x(t-1),\dots,x(0)\right] =\, \mathbb{P}\left[{x}(t)|{x}(t-1)\right], \label{eq:causal_markov}
    \end{equation}
    CVA extends this property to the approximating factors as well, $q^i_t\left(x^i(t)|{x}(t-1),\dots,{x}(0)\right)=q_t^i\left(x^i(t)|{x}(t-1)\right)$.
    Note that if additionally the observations are time-factorized, namely $\mathbb{P}[\mathcal{O}|\mathbf{x}]=\prod_{t=0}^T \mathbb{P}[\mathcal{O}_t|{x}(t)]$, then it can be shown (See Appendix I) that Markovianity extends to the posterior distribution, $\mathbb{P}[{x}({t})|{x}(t-1),\dots,{x}(0),\mathcal{O}]=\mathbb{P}[{x}(t)|{x}({t-1}),\mathcal{O}]$.
\end{enumerate}

There are simple but instructive examples where CVA leads to the exact posterior, see for example the SI epidemic model for $N=2$ individuals in Appendix A.

\Causalityacronym{} can be used to tackle some difficult problems emerging in the field of epidemic inference, such as epidemic risk assessment from partial and time-scattered observations of cases, or the detection of the sources of infection. These problems have been recently addressed within a Bayesian probabilistic framework using computational methods inspired by statistical physics \cite{baker_epidemic_2021,crisp,MCepidemics}, and generative neural networks \cite{biazzo_bayesian_2022}. With respect to existing similar approaches based on variational autoencoders (e.g. \cite{biazzo_bayesian_2022,StatMech_with_Autoreg}), the CVA ansatz for the posterior distribution does not employ neural networks, has comparatively a much smaller set of parameters and allows for much simpler physical interpretations.
In particular, risk assessment from contact tracing data is of major importance for epidemic containment, because having access to an accurate measure of the individual risk can pave the way to effective targeted quarantine plans based on contact tracing devices \cite{ferretti_contact_tracing,cencetti_contact_tracing,eames_contact_2003}. Moreover, in epidemic problems, there are quantities of interest that are not known \textit{a priori}. An example is the infection rate of the disease. Our method can be used to compute such quantities, treated as hyperparameters of the \Causalityacronym{} distribution. Being able to find the prior parameters of a distribution gives also the possibility to simplify the inference problem by adopting a simpler model. For example, in the context of epidemic inference, CVA allows one to study inference problems related to the SEIR model (introduced later) with an effective SI model, which is simpler than SEIR. This part, which we call \textit{model reduction} is illustrated in detail in the Results section.
After presenting the \Causalityacronym{} in a general setting, its main features will be discussed by exploiting a conditioned random walk \cite{conditionedRW,2DconditionedRW,geometry_conditionedRW} as a toy model. Then, an application to the important problem of epidemic inference and risk assessment on dynamic contact networks is developed and analyzed in detail. We stress that the two reference cases, i.e. the epidemic inference and the conditioned random walk, represent two very different dynamical processes, the former continuous in time while the latter  advances in discrete time-steps.

\section{Methods}

The method is based on approximating the original constrained process
by introducing an effective unconstrained causal process that is naturally
consistent with the observations.

Let $Q_{\theta}(\mathbf{x})$ be the probability distribution of a generalized dynamics, parametrized
by the vector $\theta$ of parameters. The best approximation to $\mathbb{P}\left[\mathbf{x}|\mathcal{O}\right]$
(in a precise variational sense) can be obtained by observing that
Eq.\eqref{eq:post} can be interpreted as a Boltzmann distribution
$Z^{-1}\exp\left[-H\left(\mathbf{x}\right)\right]$ with $H\left(\mathbf{x}\right)=-\log\mathbb{P}\left[\mathbf{x},\mathcal{O}\right]$
and $Z=\mathbb{P}\left[\mathcal{O}\right]$ and 
 by minimizing the corresponding variational free energy \cite{parisi_statistical_1998}, i.e.
\begin{align}
\mathcal{F}\left(Q_{\theta}\right) & :=\intop\mathrm{d}\mathbf{x}Q_{\theta}\left(\mathbf{x}\right)\log\frac{Q_{\theta}\left(\mathbf{x}\right)}{\mathbb{P}\left[\mathbf{x},\mathcal{O}\right]}\label{eq:KLdivergence}\\
 & =\left\langle \log\frac{Q_{\theta}\left(\mathbf{x}\right)}{\mathbb{P}\left[\mathbf{x},\mathcal{O}\right]}\right\rangle _{Q_{\theta}} \label{eq:free_energy}.
\end{align}
This quantity can be estimated efficiently by sampling the distribution
$Q_{\theta}$. Note that $\mathcal{F}\left(Q_{\theta}\right)=D_{KL}\left(Q_{\theta}||\mathbb{P}\left[\mathbf{x}|\mathcal{O}\right]\right)-\log\mathbb{P}\left[\mathcal{O}\right]$
where $D_{KL}$ is the Kullback-Leibler divergence \cite{kull-leib_review}. To optimize $\mathcal{F}$, 
a gradient descent method can be employed (see Appendix E), where the gradient can
be also estimated by means of sampling, indeed
\begin{align}
\nabla_{\theta}\mathcal{F}\left(Q_{\theta}\right) & =\left\langle \nabla_{\theta}\log Q_{\theta}\left(\mathbf{x}\right)\log\frac{Q_{\theta}\left(\mathbf{x}\right)}{\mathbb{P}\left[\mathbf{x},\mathcal{O}\right]}\right\rangle _{Q_{\theta}}.
\end{align}
The crucial point in this estimation is that both $Q_{\theta}$ and
$\mathbb{P}\left[\mathbf{x},\mathcal{O}\right]=\mathbb{P}\left[\mathbf{x}\right]\mathbb{P}\left[\mathcal{O}|\mathbf{x}\right]$
have explicit expressions that, due to their causal structure, can
be efficiently computed using rejection-free sampling, at difference with $\mathbb{P}\left[\mathbf{x}|\mathcal{O}\right]$ and $\mathbb{P}\left[\mathcal{O}\right]$
which do not benefit from this property. The fact that the samples are independent allows for trivial parallelism in implementation. For a detailed description of the gradient descent optimization adopted in this work, we refer the reader to Appendix E. 

When a fixed point is reached, the
corresponding distribution $Q_{\theta}\left(\mathbf{x}\right)$ is the argument that (locally) minimizes the free energy in \eqref{eq:free_energy} therefore providing an approximation of the posterior distribution $\mathbb{P}\left[\mathbf{x}|\mathcal{O}\right]$. Finally, the result can be used to generate samples satisfying the constraints given
by $\mathcal{O}$ and to compute interesting observables from them by efficiently computing sample averages.

\subsection*{A toy model application: conditioned random walk \label{sec:RW_maintext}}

\begin{figure*}
\centering
\includegraphics[width=0.9\textwidth]{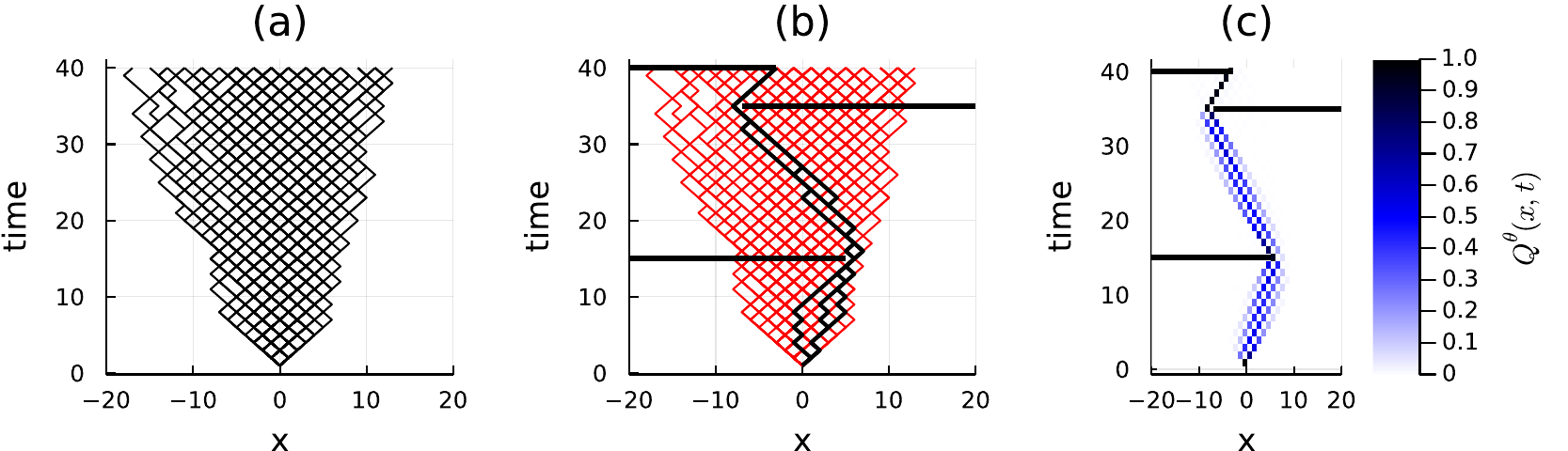}

\caption{\textit{Panel (a)} Unconditioned homogeneous random walk on a one-dimensional lattice. Time is reported on the vertical axis (up to $T=40$) and the spatial coordinate $x$ is on the horizontal axis. \textit{Panel (b)} Some trajectories are sampled from the unconditioned homogeneous distribution. The black (red) ones (do not) satisfy the constraints, i.e. they (do not) avoid the black horizontal barriers. The fraction of feasible trajectories among a given pool can be numerically estimated, and it approaches $10^{-6}$. In other words, only one of a million trajectories sampled from the unconditioned distribution satisfies the constraint. \textit{Panel (c)} The distribution of the trajectories sampled from the \Causalityacronym{} distribution. The color of each pixel indicates the probability for a trajectory to visit the corresponding state at a specific time. 
\label{fig:randomwalk}
}
\end{figure*}
Before introducing the main scenario where CVA is employed - i.e. on epidemic spreading models -, we first discuss a simple but instructive application of the method, that consists in generating an approximate probabilistic description for a conditioned random walk. A simple realization of the latter is a one-dimensional random walk, starting at site $x\left(0\right)=0$. If the generating process is spatially homogeneous,
the probability of every feasible trajectory $\mathbf{x}$ of
length $T$ is $\mathbb{P}\left[\mathbf{x}\right]=2^{-T}$. Note that every possible trajectory can be directly sampled by means of a causal generative process, namely a discrete-time Markov chain in
which the conditional probability of a jump is $\mathbb{P}\left[x\left(t+1\right)=x\left(t\right)\pm1\,|x\left(t\right)\right]=1/2$. Fig.~\ref{fig:randomwalk}(a) displays a space-time representation for a set
of realizations of such an unbiased random walk (black paths). 
For this process, let us imagine a procedure that, given a time instant $t^\mu$ and position $x^\mu$ in space, can test if the trajectory was at time $t^\mu$ to the left or right of position $x^\mu$, and denote the corresponding half-line as $W^\mu \subseteq\mathbb{Z}$. Assume that for a given unknown trajectory, we have $M$ observations of this kind $\mathcal{O}=(t^\mu,W^\mu)_\mu$ with $\mu=1,\dots,M$. The posterior
probability of a trajectory $\mathbf{x}$ can be written as
\begin{equation}
\mathbb{P}\left[\mathbf{x}|\mathcal{O}\right]=\frac{\prod_{\mu=1}^M\mathbb{I}[x\left(t^\mu\right)\in W^\mu]}{\sum_{\mathbf{y}}\prod_{\mu=1}^M\mathbb{I}[y\left(t^\mu\right)\in W^\mu]},
\label{eq:postRW}
\end{equation}
where the numerator is $1$ only if the trajectory $\mathbf{x}$ satisfies all observations, and zero otherwise. The denominator is the sum over all trajectories (so the variable $\mathbf{y}$ runs over the space of all the possible trajectories) of the numerator and plays the role of a normalization term for the posterior. The effect of $\mathcal{O}$ is to select (or constrain to) a subset of the trajectories of a free random walk, i.e. those compatible with the observations. One could naively sample trajectories from the free dynamics and then select only those compatible with $\mathcal{O}$. However, as depicted in Figure \ref{fig:randomwalk}, the fraction of trajectories compatible with the constraints might be very small to allow for a feasible computation: in the example of Fig. \ref{fig:randomwalk}(b), where it is assumed that three regions at specific time steps (black horizontal barriers) cannot be crossed, several realizations of the unconstrained dynamics are discarded (red paths), while only a small fraction is kept (black paths). In this regard, the \Causalityacronym{} provides an efficient way of generating trajectories compatible with the constraints by building up an effective probability distribution that is - by construction - compatible with the former.
Within the framework provided by \Causalityacronym{}, the following causal \textit{ansatz} can be introduced for the conditioned random walk problem:
\begin{equation}
Q_{\theta}\left(\mathbf{x}\right) =  \delta_{x\left(0\right),0}\prod_{t=0}^{T-1}\left[r_{x\left(t\right)}^{t}\delta_{x\left(t+1\right),x\left(t\right)+1} +l_{x\left(t\right)}^{t}\delta_{x\left(t+1\right),x\left(t\right)-1}\right],\label{eq:qthetarandomwalk}
\end{equation}
with $\theta=\left\{ r_{x}^{t}\right\} _{x=-T,\dots, T}^{t=1,\dots, T}$ being the set of site-dependent and time-dependent rates to jump to the right, and $l_{x(t)}^{t}=1-r_{x(t)}^{t}$ the associated probabilities to jump to the left.
We remark that Eq. \eqref{eq:qthetarandomwalk} has the same functional form as the unconstrained distribution, i.e. it still represents the probability distribution of a random walk, but with heterogeneous (in general, both in space and time) jump rates.
The distribution $Q_{\theta}$ requires $T\left(2T+1\right)$ parameters, where $2T+1$ is the total number of sites that can be visited by the realizations of the random walk. These parameters are sought by minimizing the KL distance between $Q_{\theta}$ and
the posterior distribution Eq. \eqref{eq:postRW}. The resulting probability $Q_{\theta}\left(\mathbf{x} \right)$ obtained using the \Causalityacronym{} is characterized by heterogeneous rates $r_{x}^{t}$ whose dependence in time and space perfectly mirrors the constraints introduced by the barriers. The marginal distributions of the trajectories sampled from $Q_{\theta}\left( \mathbf{x}\right)$ are represented in Fig. \ref{fig:randomwalk}(c), where the color gradient is associated with the marginal probability of occurrence of each step.

\subsection*{Epidemic Models and observations}
From now on, we will consider a class of individual-based epidemic
models describing a spreading process in a community
of $N$ individuals, interacting through a (possibly dynamic) contact network.
The overall state of the system at time $t$ (consisting of the state
of each individual) is described by a vector $\mathbf{x}\left(t\right)\in\mathcal{X}^{N}$,
where $\mathcal{X}$ is a finite set of possible health conditions
(called compartments). The simplest, but already non-trivial, model
of epidemic spreading is the discrete-time Susceptible-Infected (SI)
model \cite{SI_description_1994}, in which $\mathcal{X}=\left\{ S,I\right\} $ (corresponding to an individual being ``susceptible'' and ``infected'', respectively)
where the only allowed transition occurs from state $S$ to state
$I$. More precisely, each time $t$, if an infected
individual $j$ is in contact with a susceptible individual $i$, the former can
infect the latter (which moves into state $I$) with a transmission probability $\tilde{\lambda}_{ji}\left(t\right)$, sometimes called
\emph{transmissivity}.  Since transmissions are independent, the
individual transition probabilities are
\begin{equation}
\mathbb{P}\left[x_{i}\left(t+\Delta t\right)=S|{x}\left(t\right)\right]=\delta_{x_{i}\left(t\right),S}\prod_{j\neq i}\left(1-\tilde{\lambda}_{ji}\left(t\right)\delta_{x_{j}\left(t\right),I}\right)\label{eq:SI1}
\end{equation}
and $\mathbb{P}\left[x_{i}\left(t+\Delta t\right)=I|{x}\left(t\right)\right]=1-\mathbb{P}\left[x_{i}\left(t+\Delta t\right)=S|{x}\left(t\right)\right]$.
A simple assumption is that time dependence only enters to describe
the dynamic nature of the contact network (with $\tilde{\lambda}_{ji}\left(t\right)=0$
if there is no contact between $j$ and $i$ at epoch $t$). More
realistically, the transmission probability $\tilde{\lambda}_{ji}\left(t\right)$
should also depend on the current stage of infection of the infector
$j$ (e.g. presence/absence of symptoms) and thus mainly on the time
elapsed since her own infection, making the epidemic dynamics non-Markovian.

Assuming transmission probabilities proportional to $\Delta t$ in
Eq.\eqref{eq:SI1} and defining transmission rates as $\lambda_{ij}\left(t\right)=\lim_{\Delta t\to0^{+}}\tilde{\lambda}_{ij}\left(t\right)/\Delta t$,
a continuous-time version of the SI model is obtained. Both for the
discrete-time and continuous-time SI models, the history of the epidemic
process can be fully specified by the ``infection times'' $\left\{ t_{i}\right\} _{i=1}^{N}$
of all individuals; we conventionally set
$\ensuremath{t_{i}=0}$ if individual $i$ is already infected at
the initial time and $t_{i}=+\infty$ if $i$ is never infected during
the whole epidemic process. In terms of the trajectory
vector $\mathbf{t}:=\left(t_{1},\dots,t_{N}\right)$, the probability pseudo-density of an epidemic history can be generally written as 
\begin{align}
\mathbb{P}\left[\mathbf{t}\right] & = \prod_{i}\left\{ \gamma\delta\left(t_{i}\right)+\left(1-\gamma\right)\Lambda\left(\sum_{j\neq i}\mathbb{I}\left[t_{j}\leq t\right]\lambda_{ji}\left(t\right),t_{i}\right)\right\},\label{eq:Intro:SI_continuous-2}
\end{align}
where $\gamma$ denotes the probability of each individual to be a patient zero, and  $\Lambda\left(f(t),b\right)=f\left(b\right)e^{-\int_{-\infty}^{b}f\left(t\right)dt}$
is the first-success distribution density of an event
with rate $f(t)$. Hence, the quantity $\Lambda\left(\sum_{j\neq i}\mathbb{I}\left[t_{j}\leq t\right]\lambda_{ji}\left(t\right),t_{i}\right)$ is the probability density associated with the infection, at time $t_{i}$,
of individual $i$ by one of its infectious contacts at previous times. 
Notice that when $\int_{-\infty}^{b}\Lambda\left(f,t\right)dt<1$, it means that
there is a non-zero probability of the individual remaining susceptible.
In that case, we will formally assign the defect mass $1-\int_{-\infty}^{b}\Lambda\left(f,t\right)dt$ 
to $t=\infty$.
In addition to the epidemic model, a set of observations has to be defined. 
In real epidemics, observations mirror the outcomes of medical tests, namely the state of an individual $i$ at time $t$. For the sake of simplicity, an auxiliary variable $r \in \{+. -\}$ representing positive or negative tests, respectively, is defined, such that each observation can be encoded as a triplet $\left(i,t,r\right)$. 
Given the
stochastic nature of clinical tests, it is assumed that the outcome $r$
of a test performed on individual $i$ at time $t$ obeys
a known conditional distribution law $\mathbb{P}\left[r|{t}_{i}\right]$
where ${t}_{i}$ represents the infection time of individual $i$. When medical tests are affected by uncertainty, i.e. there exist non-zero false positive and false negative rates of the diagnostic tests, the conditional probability states
\begin{subequations}
\begin{align}
\mathbb{P}\left[r=+|{t}_{i}\right] & =\left(1-p_{\rm{FNR}}\right)\,\mathbb{I}\left[t_{i}\leq t\right]+p_{\rm{FPR}}\,\mathbb{I}\left[t_{i}>t\right]\\
\mathbb{P}\left[r=-|{t}_{i}\right] & =p_{\rm{FNR}}\,\mathbb{I}\left[t_{i}\leq t\right]+\left(1-p_{\rm{FPR}}\right)\,\mathbb{I}\left[t_{i} >t\right]
\end{align}
\end{subequations}

For a population undergoing $M$ individual test events,
the set $\mathcal{O}$ of observations is then identified with the
set of triplets $\left(i^{\mu},t^{\mu},r^{\mu}\right)$ for $\mu=1,\dots,M$.
As in the random walk example, each observation constrains the dynamics: in the noise-less case (i.e. $p_{\rm{FNR}}=p_{\rm{FPR}}=0$) the posterior distribution gives zero measure to all the epidemic trajectories that violate the observations.
Given a realization of an epidemic model defined on a contact network, and the (possibly noisy) observation of the states of a subset of the individuals (at
possibly different times), the epidemic risk assessment problem consists of estimating
the epidemic risk, i.e. the risk of being infected, of the unobserved
individuals at some specific time. In practice, it amounts to computing
marginal probabilities from the posterior distribution
\begin{equation}
\mathbb{P}\left[x_{i}\left(t\right)=I|\mathcal{O}\right]=\int d\mathbf{t}\mathbb{I}\left[t_{i}\leq t\right]\mathbb{P}\left[\mathbf{t}|\mathcal{O}\right] \label{eq:riskSI}
\end{equation}
where $\int d\mathbf{t}$ denotes the integral over all infections
times $t_{1},\dots,t_{N}$. 

A richer epidemic model, that is often used as a testing ground for more realistic scenarios (see e.g. Refs \cite{kerr2021covasim,hinch2021openabm}), is the SEIR model \cite{seir_2014}, which includes also the Exposed (E) and Recovered (R) states, i.e. $\mathcal{X}=\left\{ S,E,I,R\right\} $. The only allowed transitions are $S\to E$ (representing the contagion event), $E\to I$, and $I\to R$; the latter ones occur for each individual independently of the others, with latency and recovery rates $\nu_i$ and $\mu_i$, respectively. 
The previous representation in terms of transmission times can be straightforwardly generalized to the SEIR model (by introducing individual-wise infective and recovery times) as well as the definition of observations from clinical tests and the measure of the individual risk (see Appendix B for additional details). 

The choice of the parameters $\theta$ of the \Causalityacronym{} \textit{ansatz} reflects and somehow generalizes the features of the generative model $\mathbb{P}\left[ \mathbf{t} \right]$: in the SI case for instance, for each individual $i$, heterogeneous infection rates $\lambda_{i}\left(t\right)$, zero-patient probabilities $\gamma_{i}$, and self-infection rates $\omega_{i}(t)$ are defined. Since  $\lambda_{i}\left(t\right)$ and $\omega_{i}\left(t\right)$ are time-dependent quantities, an additional parametrization is introduced for computational purposes. Then, the trial distribution $Q_{\theta}\left(\mathbf{t}\right)$ is optimized with respect to the full set of parameters. The total number of parameters for the inference in the SI model in a population of $N$ individuals is $7N$, while for the SEIR model is $13N$.
We refer to Appendix C for a more detailed discussion of the parameter choice and of the implementation of the gradient descent.
\section{\label{sec:EpidemicRiskAss}Results on epidemic inference}

\subsection*{Performances on synthetic networks}

The performance of the \Causalityacronym{} in reconstructing epidemic trajectories can be tested by measuring its ability in classifying the state of the unobserved individuals based on their predicted risk. The results are directly compared with those obtained using other inference techniques previously proposed in the literature, such as Belief Propagation \cite{mezard_information_2009} as implemented in   \cite{baker_epidemic_2021} (sib), a Monte Carlo method (MC), a Soft Margin method (soft) adapted from \cite{softmarg}, and simple heuristic methods based on Mean Field equations (MF) \cite{baker_epidemic_2021} or on sampling (heu). A description of the implementation of the several methods used for comparison is provided in the Appendix G.

\begin{figure*}[tb]
\centering{}\includegraphics[width=0.6\textwidth]{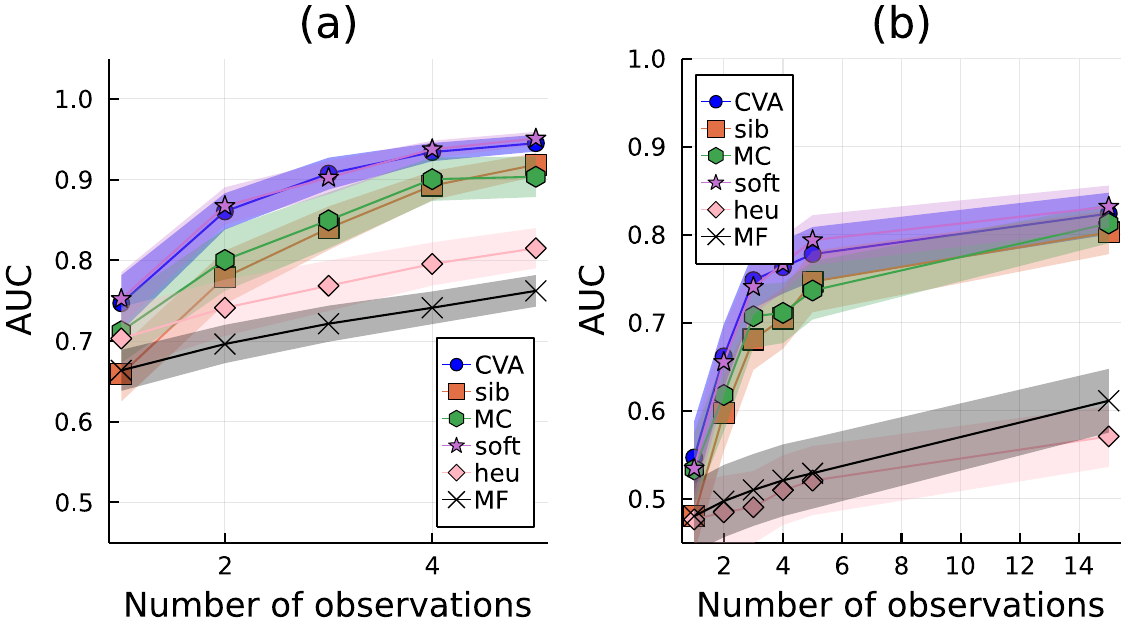}
\caption{Area under the ROC (AUC) as a function of the number of observations for the risk assessment problem, i.e. $t^{\star} = T$, in panel (a), and for the patient-zero problem, $t^{\star} = 0$, panel (b). 
The simulated contact graph is a proximity network with average connectivity $2.2/N$. For both simulations in panels (a) and (b), the total number of individuals is $N=50$, the probability of being the zero patient is set to $\gamma=1/N$, and the infection rate is $\lambda=0.1$. For each epidemic realization, the inference is performed for an increasing number of noiseless observations (here $p_{\rm{FNR}}=0$) at time $t_{\rm{obs}} = T$. 
Thick lines and shaded areas indicate the averages and the standard errors computed over $40$ different instances.
\label{fig:The-zero-patient}} 
\end{figure*}

For the sake of simplicity,
we considered SI epidemic processes on proximity
graphs \cite{proximity_2019}, i.e. random graphs generated by proximity relationships
between $N = 50$ individuals randomly drawn from a uniform distribution on a two-dimensional square region (a definition of proximity graphs is given in Appendix H). Each instance corresponds to a different realization of both the dynamical network and the forward epidemic propagation.   Observations $\mathcal{O}$ are noiseless, i.e. $p_{\rm{FNR}} = 0$, and performed on a randomly chosen fraction of the population at
a fixed time $t_{{\rm obs}} = T$. The comparison among the different inference methods
is performed by ranking the individual marginal probabilities of being in
state $I$ at a chosen time $t^{*}$ and building the corresponding \emph{receiving operating characteristic} (ROC) curve \cite{fawcett_introduction_2006}. The AUC (area under the ROC curve) at a time $t^{*}$ is an indicator
of the accuracy of the method in reconstructing the state of the individuals
at that time. The AUC at initial time ($t^{*}=0$, patient-zero problem)
and at the final time ($t^{*}=T$, risk assessment) are shown in Figure
\ref{fig:The-zero-patient} as function of the  number of observations available. 

As one may expect, in both cases the average performances of all methods improve when the number of observations increases. In particular, the Soft Margin method is expected to converge to the exact results for this type of experiment when the number $N$ of individuals is small. The results obtained with \Causalityacronym{} are very close (and closer than any other technique) to those obtained by means of Soft Margin (soft), even in the interesting and challenging regime with only a few observations.

To further investigate the performances of \Causalityacronym{} against other state-of-the-art techniques, the AUC associated with the prediction of individual risk is quantified as a function of time, with two different observation protocols: (i) the states of a fraction of individuals are observed at observation times scattered over the duration of the simulation or (ii) observations are performed at the last time $t_{\rm{obs}} = T$. More precisely, in the first protocol observation times are randomly drawn \textit{a priori} with uniform distribution  in the interval $\left[1,T\right]$, and observations are biased towards tested-positive outcomes to mimic a realistic scenario where symptomatic, i.e. infected individuals, are more likely tested than susceptible ones.  For these experiments, two realistic dynamic contact network instances are considered, one generated using the Spatio-temporal Epidemic Model (StEM) in continuous-time in Ref.~\cite{lorch2022quantifying} and the other using the discrete-time OpenABM model in Ref. ~\cite{ferretti_contact_tracing} (see Appendix H for a brief description of StEM and OpenABM models).
For sake of simplicity, instead of adopting the complex epidemic dynamics described in Ref. ~\cite{ferretti_contact_tracing} and Ref.~\cite{lorch2022quantifying}, epidemic realizations are generated using a continuous-time SI model on these contact graphs.
 
\begin{figure*}
\centering{}\includegraphics[width=0.8\textwidth]{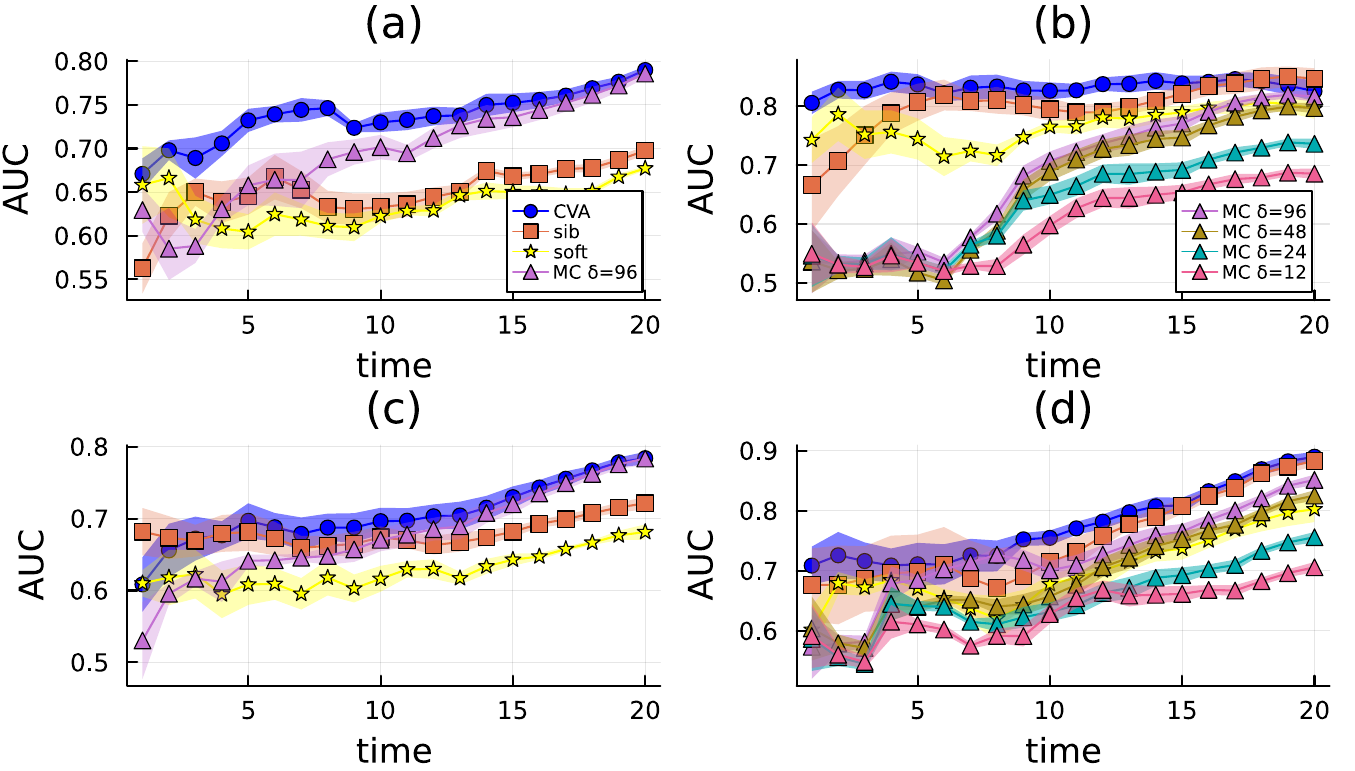}\caption{AUC associated with the prediction of the infected individuals, for the \Causalityname{} (CVA), Belief Propagation (sib) and SoftMargin (soft), and MCMC (MC) as a function of time during the epidemic propagation of a SI model on several instances of dynamic contact network generated using the OpenABM model  ~\cite{ferretti_contact_tracing} (in panel (a) $N = 2000$, in (c) $N = 1000$) and the StEM in Ref.~\cite{lorch2022quantifying} (panels (b) and (d)) for $N = 904$. The infection rate is set to $\lambda=0.15$ for the latter and $\lambda = 0.02$ for the former; observations are noiseless in both cases.
For panels (c) and (d), observations are performed at the last time of the dynamics, i.e. $t_{\rm{obs}} = T$. For the results in panels (a) and (b) observation times are extracted uniformly in the range $\left[1, T \right]$; at each observation time $t_{\rm obs}$, infected nodes are observed with a biased probability equal to $1.1\times  N_{I}\left(t_
{\rm obs}\right)/N$ where $N_{I}\left(t_{\rm obs}\right)$ is the number of infected individuals at time $t_{\rm obs}$ and $N$ is the total number of individuals. The total number of observations is $n_{\rm{obs}} = N \cdot 0.1$ for OpenABM and $n_{\rm{obs}} = 100$ for the StEM.
\label{fig:Tubingen}} 
\end{figure*}

A measure of the individual risk is computed according to all different methods (CVA, sib, soft, and MC), and the corresponding AUCs are shown as functions of time (in days), in Figure \ref{fig:Tubingen}(a,c) and \ref{fig:Tubingen}(b,d) for OpenABM and the StEM respectively. For the latter only, we also consider different MC parameterizations, in particular when using $\delta \in \{12, 24, 48, 96\}$ hours; a further increase of $\delta$ does not carry any improvement of the results. The quantity $\delta$ is associated with the MC move's proposal (see Appendix G for a detailed description). Panels (a) and (b) are associated with the observations scattered in time, while panels (c) and (d) use observations at the last time only. In panel (a) simulations are run for $N = 2000$, while in panel (c) we set $N = 1000$.
It is easy to see that, in panels (a) and (c), CVA (blue dots) is the best-performing method in terms of AUC; only MC (pink triangles) reaches comparable AUC for $t \sim T$. The results achieved by Belief Propagation are similar to those produced by CVA when the size of the graph is $N = 1000$ (panel (c)), and significantly deteriorate for $N = 2000$ (panel (a)).
For the instances generated according to StEM in panels (b) and (d), the comparison reveals that CVA achieves the largest values of the AUC at all times and only Belief Propagation (sib, orange squares) performs comparably to CVA for the risk assessment problem, i.e. the inference at the last time of the dynamics. MC for $\delta \in \{48, 96 \}$, approaches CVA performances in the last days while is not able to predict the zero patient. Indeed, the AUC associated with MC predictions for all parametrizations is slightly larger than 0.5 for $t < 5$ when observations are performed at the last time of the dynamics. 

\subsection*{Hyperparameters inference}
\begin{figure*}
\centering{}\includegraphics[width=0.8\textwidth]{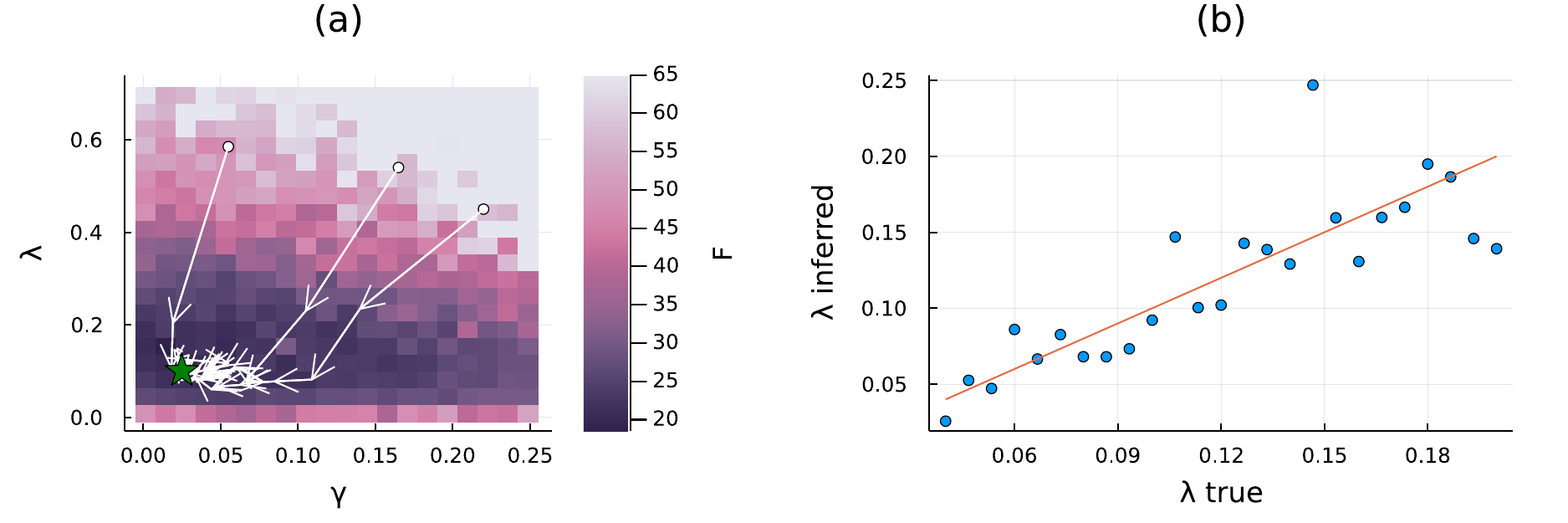}\caption{\label{fig:Results:Hyperparams} \textit{Panel (a)} Heat map of the free energy ($F := -\log \mathbb{P}(\mathcal{O})$) computed at the convergence of \Causalityacronym{} as a function of the assumed hyperparameters of the generative SI model. The experiment is performed on a proximity graph with $N=50$ individuals and density $\rho=2/N$; the epidemic model is characterized by the zero-patient probability $\gamma^*=1/N$ and the infection rate $\lambda^*=0.1$, shown here as a green star. We perform a large number of observations ($n_{\rm{obs}}=2N$) at uniformly randomly distributed times. As expected, the lowest values of this free energy are concentrated around the exact value $(\gamma^*,\lambda^*)$.
The oriented paths (white arrows) represent the convergence towards the minimum of $-\log \mathbb{P}[\mathcal{O}]$ obtained by performing a gradient descent algorithm over the hyperparameters starting from three different initial points in the plane $(\gamma,\lambda)$. \textit{Panel (b)} Scatter plot of inferred values for the infection probability against the ground truth. In these experiments, we fix and assume to know the zero patient probability $\gamma=1/N$ while the infection parameter $\lambda$ is varied. For each $\lambda$ an epidemic simulation is performed and $n_{\rm{obs}}=10N$ observations are taken at uniformly randomly distributed times. }
\end{figure*}

In the previous numerical experiments, the
parameters of the generative SI model, i.e. the (homogeneous) infection rate $\lambda$
and the probability of being the zero patient $\gamma$, are assumed to be known. These quantities enter the \Causalityacronym{} formalism as hyperparameters of the prior distribution which are often inaccessible in realistic applications, but can be estimated as those realizing the minimum of the free-energy $F=-\log\mathbb{P}[\mathcal{O}]$. This can be achieved by gradient descent if the number $n_{\rm obs}$ of available observations is sufficiently large (see Appendix F for details). An example of the quality of the parameter inference is provided by the following experiment. For a SI model with $n_{\rm{obs}}$ and true parameters $\left(\gamma^{\star}, \lambda^{\star}\right) = \left( 1/N, 0.1\right)$ (see the caption of Figure \ref{fig:Results:Hyperparams} for further details), 
Figure \ref{fig:Results:Hyperparams}(a) shows a heatmap of $F$ computed at the convergence of \Causalityacronym{} as function of the pair of values $\left(\gamma,\lambda\right)$ used as the hyperparameters of the corresponding prior distribution. The region attaining the lowest
values also contains the true values $(\gamma^{*},\lambda^{*})$.
The oriented paths (white arrows) in Figure \ref{fig:Results:Hyperparams}(a) represent the sequences of intermediate values of $\lambda$ and $\gamma$ obtained during the convergence process of  \Causalityacronym{}, starting from three different initial conditions.
These traces show that trajectories end up in the same region, very close to where the true values are located (green star).
Similar experiments, where the value of the zero patient probability is set to $\gamma = 1 / N$ and  the infection probability varies in the range $\left[0.05, 0.20\right]$ are performed. Similarly to the previous set-up, \Causalityacronym{} is applied to infer the parameter $\lambda$. Figure \ref{fig:Results:Hyperparams}(b) displays a scatter plot of the inferred values against the true ones. Results suggest a good agreement between the result of the inference and the generative process. 

\subsection*{Model reduction}

\begin{figure*}
\centering{}\includegraphics[width=0.8\textwidth]{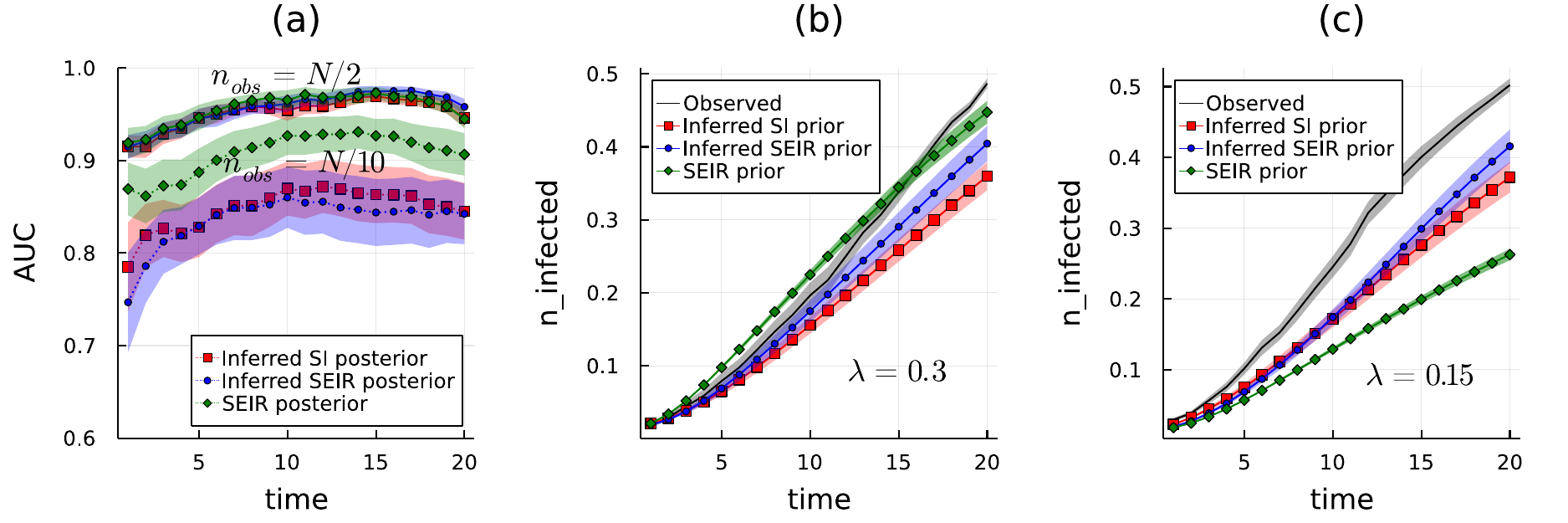}\caption{\label{fig:Results:SEIRvsSI}Effects of model reduction on inferential performances and generative capabilities. The numerical experiments are performed on a proximity graph with $N=100$ individuals and density $2.2/N$. The observed epidemic realizations are generated using an SEIR model with $\gamma=1/N$, $\lambda=0.3$ (panels (a) and (b)) and $0.15$ (panel (c)), latency delay $\nu=0.5$ and recovery delay $\mu=0.1$. 
\textit{Panel (a)} Values of the AUC as a function of time obtained using the \Causalityacronym{} in two observation regimes (when the number of observations is $n_{obs}=N/10$ and  $n_{obs}=N/2$), with the three different inferred posterior distributions:  
an SEIR model with known hyperparameters (green diamonds), an SEIR model with unknown hyperparameters (blue circles), and a SI model with unknown hyperparameters (red squares). Shaded areas represent the error around the average value, computed using $22$ instances.
\textit{Panels (b) and (c)} The average fraction of infected individuals as a function of time estimated using the correct SEIR prior model (green diamonds), an SEIR prior with the inferred hyperparameters (blue circles), and a SI prior model with the inferred hyperparameters (red squares). The regimes shown correspond to unbiased observations (center, for $\lambda=0.3$), and to observations preferentially sampled from large outbreaks (right, for $\lambda=0.15$). The black curves represent the same quantity computed from the observed epidemic realizations. Shaded areas represent the standard error computed from 40 realizations of the dynamics.
}
\end{figure*}

For viral diseases with sufficiently known transmission mechanisms,
agent-based modeling using discrete-state stochastic processes has
proven useful to build large-scale simulators of epidemic outbreaks
and design containment strategies \cite{hinch2021openabm, lorch2022quantifying}. Such mathematical models are much more complex than the SI model analyzed previously, as they need to include additional specific features of real-world diseases. In particular, models may assume different infected states, characterized by a different capability of transmitting the virus and diverse sensitivity to diagnostic tests. Another important feature that can emerge from realistic transmissions is that individuals can stop being infectious, even before recovering from the infected
state, because of the decay of their viral load. These ingredients may be effectively included in the SI and SEIR
models by introducing time-dependent infection rates, which is a natural assumption in the framework of the \Causalityname{} (see App. B-C). This property makes the latter a very suitable inference method to approximate unknown and possibly complex generative epidemic processes using classes of simpler probabilistic models.  
A simple test of such a potentiality is provided by the following example. Several epidemic realizations are generated with an SEIR prior model and the quality of the inference obtained by the \Causalityname{} is evaluated when (i) the SEIR model is also used as an \textit{ansatz} for the posterior distribution and (ii) when the posterior distribution is approximated with a simpler probabilistic model, such as the SI model.  
If the parameters of the generative SEIR model are known, the hyperparameters of the SEIR posterior are also known. The corresponding results (green diamonds) for the AUC as a function of time on a proximity graph are displayed in Fig.~\ref{fig:Results:SEIRvsSI}(a). Otherwise, the hyperparameters of the SEIR posterior can be inferred by means of the \Causalityacronym{} (blue circles). Finally, the \textit{ansatz} for the posterior distribution can be simplified to a SI model, and the corresponding hyperparameters can be inferred as well within the \Causalityacronym{} (red squares). 
The overall quality of the inference depends on the possibly different regimes of information contained in the observations. 
Strikingly, when the generative model is not known, the results from SEIR-based and SI-based inference are always very close to each other. For a sufficiently large number of observations, such results are also close to those obtained with the SEIR posterior and known hyperparameters. 

From a generative perspective, the inferred hyperparameters can be also interpreted as the epidemic parameters of some prior model, from which epidemic realizations can be sampled. It is natural to ask what are the statistical properties of such generative processes compared to the original one, from which the observations were sampled. Figure~\ref{fig:Results:SEIRvsSI} also shows, in two different regimes, as a function of time the average number of infected individuals estimated from the original SEIR prior model (green diamond), the SEIR prior model with inferred hyperparameters (blue circles) and the SI prior model with inferred hyperparameters (red squares). The average number of infected individuals computed over the realizations from which the observations are sampled is also displayed (black line).
The regimes shown in Fig.~\ref{fig:Results:SEIRvsSI} correspond to unbiased observations (panel (b), for $\lambda=0.3$), and to observations preferentially sampled from large outbreaks (panel (c), for $\lambda=0.15$). Although the discrepancy between the different curves is significant, the moderate difference between predictions obtained using SEIR and SI prior models with inferred hyperparameters suggests that model reduction is only a minor source of information bias. 

\section{Conclusions}

Sampling from the posterior distribution of a conditioned dynamical process can be computationally hard. 
In this work, a novel computational method to accomplish this task, called the \Causalityname{}, was put forward. 
The \Causalityname{} is based on the idea of inferring the posterior with an effective, unconditioned, dynamical model, whose parameters can be learned by minimizing a corresponding free energy functional.
An insight into the potential of the method is obtained by analyzing a one-dimensional conditioned random walk in which some regions of space are forbidden. The \Causalityacronym{} produces a generalized random walk process, with space-dependent and time-dependent jump rates, whose unconditioned realizations satisfy the imposed constraints. An application of greater practical interest concerns epidemic inference, in particular the risk assessment from partial and time-scattered observations. For simple stochastic epidemic models, such as SI and SEIR, taking place on contact networks of moderately small size, the \Causalityacronym{} performs better or as well as the best methods currently available. Moreover, the variational nature of the method allows one to estimate the parameters of the original epidemic model that generated the observations, which enter the \Causalityacronym{} in the form of hyperparameters. Since the \Causalityacronym{} approximates the posterior distribution of the epidemic process by learning a set of generalized individual-based, time-dependent parameters, even with a rather simple \textit{ansatz} for the epidemic model, such as an SI model, inference from observations coming from more complex epidemic processes can be performed. In fact, a generalized SI model with time-dependent infection rates and self-infection rates allows one to accommodate many features of real-world epidemic diseases, such as time-varying viral load and transmissivity, incubation, and recovery. The performances of the \Causalityname{} do not seem to suffer from model reduction from SEIR to SI, suggesting that simplified epidemic models could be effective for inference also in real-world cases.  
The \Causalityname{} is very flexible and, employing sampling to perform estimates, it can be applied virtually to any dynamics for which the latter can be carried out efficiently. In particular, the method can thus be applied to inference problems involving recurrent epidemic processes, such as the SIS model \cite{sis_lage} or other models (e.g. \cite{westnile}) where immunity decays over time. There are, however, some limitations. The \Causalityacronym{} relies on the fact that the functional form of the posterior should be similar to the one of the prior. 
This is not true in general. For example, let us take an epidemic SI model in which the zero-patient probability $\gamma$ is infinitesimally small. If one individual is tested positive at a certain time, then the posterior distribution is substantially different from the prior. In particular, in the prior process each individual is the patient zero independently with probability $\gamma$, while in the posterior there is a strong (anti)correlation: indeed the measure will concentrate on trajectories with exactly one infected individual (and this is impossible to reproduce with independent patient zero probabilities). Of course, this example is extremely contrived, as the probability that infection occurs at all in this system, and thus such a test result can be obtained, is infinitesimally small as well. Moreover, in this case the problem can be simply solved by adopting a more natural distribution for the initial state (either by using a non-infinitesimal initial infection probability in the prior, or by adopting a single initial infection in the test distribution $q$, see also Appendix D ). Nevertheless, it is a simple example in which the prior functional form is substantially different from the one of the posterior. 
\section*{Data Availability Statement}
All data is generated using simulations and can be reproduced by following the prescriptions provided in the main text and in Supplementary Information. A public GitHub repository containing a Julia implementation of the algorithm and notebooks to reproduce the results of this work is available at 
\url{https://github.com/abraunst/Causality.git}

\section*{Acknowledgement}
This study was carried out within the FAIR - Future Artificial Intelligence Research and received funding from the European Union Next-GenerationEU (Piano Nazionale Di Ripresa e Resilienza (PNRR) – Missione 4 Componente 2, Investimento 1.3 – D.D. 1555 11/10/2022, PE00000013). This manuscript reflects only the authors’ views and opinions, neither the European Union nor the European Commission can be considered responsible for them.\\
LDA acknowledges financial support from ICSC (Centro Nazionale di Ricerca in High-Performance Computing, Big Data, and Quantum Computing) founded by European Union – NextGenerationEU.

\bibliographystyle{unsrt}

\newpage

\appendix

\section{Exact posterior distribution of a SI model with two individuals and one observation\label{sec:appendixSI2nodes}}

\begin{figure*}[b]
\centering{}\includegraphics[width=0.6\textwidth]{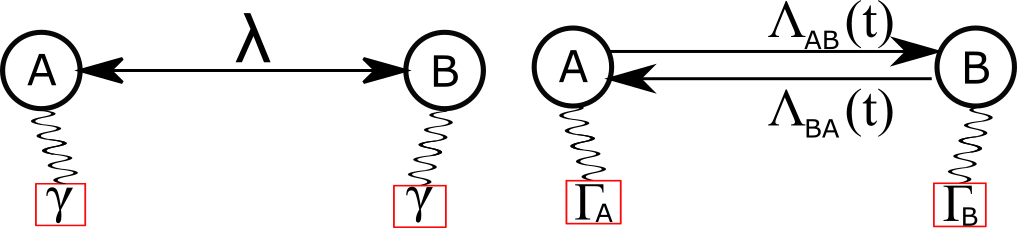}
\caption{Schematic representation of the role played by the epidemic parameters in a SI model with two individuals. \textit{Left}: In the original SI model, individuals A and B have the same probability $\gamma$ of being initially infected, and the same infection rate $\lambda$. \textit{Right}: in the prior SI model used to compute the posterior by means of the \Causalityacronym{}, $\gamma$ are replaced by possibly different probabilities $\Gamma_{A},\Gamma_{B}$ and the infection rate $\lambda$ is replaced by time-dependent rates $\Lambda_{\text{A}\text{B}} (t),\Lambda_{\text{B}\text{A}} (t)$. These parameters are inferred by minimizing the KL divergence in Eq. \eqref{eq:KLdivergence} in the main text.  \label{fig:SIn=2}}
\end{figure*}

A simple but instructive example to understand how the \Causalityname{} can obtain good approximations of the posterior distribution in epidemic processes is provided by analyzing the case of a continuous-time SI model  
with only two individuals, A and B. In this example, we assume that (i) both individuals have the same probability $\gamma$ of being initially infected and (ii) they can infect each other, if infectious, with the same constant infection rate $\lambda$. A graphical representation of the role played by these parameters is shown in Figure~\ref{fig:SIn=2} (\textit{left}).

Suppose an individual A is observed at time $t_{\rm obs}=T$ in the infected state (e.g., performing a clinical test), then there are two possible explanations for such an observation: either A was already infected at time $t=0$
or A got infected by B at a time $t<T$. In both cases, the observation
forces the dynamics to satisfy the constraint that
A is in state $I$ at time $T$. One can speculate that the observation breaks the causality property of the dynamics because a constraint set on the future (time
$T$) affects the dynamics at previous times. If, in fact, A is not the zero patient, then B must have infected A at previous
times. The effective posterior infection rate from B to A, therefore, must diverge when $t\to T$.

The epidemic trajectory can be described defining $\boldsymbol{t}:=(t_{\text{A}},t_{\text{B}})$, where $t_\text{A}$ and $t_{\text{B}}$ are the infection times of the two individuals. In terms of $\boldsymbol{t}$, the SI prior distribution can be written  as
\begin{align}
\mathbb{P}[\boldsymbol{t}] & =\begin{cases}
\gamma^{2} & t_{\text{A}}=t_{\text{B}}=0,\\
(1-\gamma)^{2} & t_{\text{A}}=t_{\text{B}}=\infty\text{,}\\
\gamma(1-\gamma)e^{-\lambda t}\lambda & t_{\text{A}}=t\text{ and \ensuremath{t_{\text{B}}=0}\ensuremath{\,};\ensuremath{\,t_{\text{A}}=0} and \ensuremath{t_{\text{B}}=t} },\\
0 & \text{otherwise.}
\end{cases}
\end{align}
This prior probability distribution 
\footnote{There is a little abuse of notation. $p_{\text{A}\text{B}}(0,0)$ and $p_{\text{A}\text{B}}(\infty,\infty)$
are probabilities, while the other term is a density of probability. The point is that the event of having two zero patients or no infections
have finite probability, while the event of having a particular infection
time is infinitesimal in probability and must be integrated over time.} 
is normalized, indeed
\begin{equation}
\gamma^{2}+2\gamma(1-\gamma)\lambda\intop_{0}^{\infty}e^{-\lambda t}dt+(1-\gamma)^{2} =\gamma^{2}+2\gamma(1-\gamma)+(1-\gamma)^{2}=1.
\end{equation}
The observation of the state of individual A implies that 
\begin{align}
\mathbb{P}[\mathcal{O}|\boldsymbol{t}] =\begin{cases}
0 & \text{if }t_{\text{A}}>T\\
1 & \text{otherwise}.
\end{cases}
\end{align}
The posterior distribution $\mathbb{P}[\boldsymbol{t}|\mathcal{O}]$ can be computed using Bayes theorem as
\begin{align}
\mathbb{P}[\boldsymbol{t}|\mathcal{O}]  &=\frac{\mathbb{P}[\boldsymbol{t}]\mathbb{P}[\mathcal{O}|\boldsymbol{t}]}{\mathbb{P}[\mathcal{O}]}\\
& = \frac{1}{\mathbb{P}[\mathcal{O}]}\begin{cases}
\gamma^{2} & t_{\text{A}}=t_{\text{B}}=0\\
\gamma(1-\gamma)e^{-\lambda t_{\text{A}}}\lambda & 0<t_{\text{A}}<T\text{ and \ensuremath{t_{\text{B}}=0}}\\
\gamma(1-\gamma)e^{-\lambda t_{\text{B}}}\lambda & t_{\text{A}}=0\\
0 & \text{otherwise.}
\end{cases}\label{eq:Intro:N=00003D2posterior}
\end{align}
where the denominator is given by 
\begin{align}
\nonumber \mathbb{P}[\mathcal{O}] & =\gamma^{2}+\int_{0}^{T}dt_{\text{A}}\,p_{\text{A}\text{B}}(t_{\text{A}},0) + \int_{0}^{\infty}dt_{\text{B}}\,p_{\text{A}\text{B}}(0,t_{\text{B}})\\
 & =\gamma^{2}+\gamma(1-\gamma)\left(2-e^{-\lambda T}\right).
\end{align}

It is convenient to compute the posterior probability  that individuals are infected at the initial time given the observation $\mathcal{O}$, 
\begin{align}
\mathbb{P}[t_{\text{A}}=0 |\mathcal{O}] & =\frac{\gamma}{\gamma^{2}+\gamma(1-\gamma)\left(2-e^{-\lambda T}\right)}=:\Gamma_{\text{A}}\label{eq:Intro:N=00003D2posterior_seedA}\\
\mathbb{P}[t_{\text{B}}=0 |\mathcal{O}]& =\frac{\gamma+\left(1-\gamma\right)\left(1-e^{-\lambda T}\right)}{\gamma+(1-\gamma)\left(2-e^{-\lambda T}\right)}=:\Gamma_{\text{B}}\label{eq:Intro:N=00003D2posterior_seedB}
\end{align}
For both individuals, non-causal effects arise as the infection probabilities at time 0 also depend on the infection rate $\lambda$ and on the 
observation time $T$. Moreover, the expressions of $\Gamma_{\text{A}},\Gamma_{\text{B}}$ are different because the observation on A has broken the symmetry between the two individuals. 

It is also possible to compute the posterior probability that individual B is infected at time $t$, which implies that A is initially infected, namely
\begin{align}
\nonumber\Lambda_{\text{A}\text{B}}(t) & :=\lim_{\epsilon \to 0}\frac{\mathbb{P}[t_{\text{B}}=t+\epsilon,t_{\text{A}}=0|\mathcal{O}]}{\mathbb{P}[t_{\text{B}}>t,t_{\text{A}}=0|\mathcal{O}]}\\
&\, =\lim_{\epsilon \to 0} \frac{\gamma(1-\gamma)e^{-\lambda (t+\epsilon)}\lambda}{\gamma(1-\gamma)e^{-\lambda t}}=\lambda \label{eq:Intro:N=00003D2postInteractionAB}
\end{align}
and the posterior probability that individual A is infected at time $t$ by B, i.e.
\begin{align}
\nonumber \Lambda_{\text{B}\text{A}}(t) & :=\lim_{\epsilon \to 0} \frac{\mathbb{P}[t_{\text{A}}=t+\epsilon,t_{\text{B}}=0|\mathcal{O}]}{\mathbb{P}[t_{\text{A}}>t,t_{\text{B}}=0|\mathcal{O}]}\\
& \nonumber = \lim_{\epsilon \to 0} \frac{\gamma(1-\gamma)e^{-\lambda (t+\epsilon)} \lambda}{\gamma(1-\gamma)\lambda\intop_{t}^{T}e^{-\lambda s}ds}=\nonumber \\
 & =  \lim_{\epsilon \to 0} \frac{e^{-\lambda( t+\epsilon)} \lambda}{\left(e^{-\lambda t}-e^{-\lambda T}\right)}= \frac{\lambda e^{-\lambda t}  }{e^{-\lambda t}-e^{-\lambda T}}. \label{eq:Intro:N=00003D2postInteractionBA}
\end{align}
The two quantities differ because here the observation forces individual B to infect A before the observation time $t_{\rm obs}=T$, causing the infection rate $\Lambda_{\text{B}\text{A}}(t)$ to diverge when $t \to T$.

\begin{figure*}[tb]
\centering{}\includegraphics[width=0.6\textwidth]{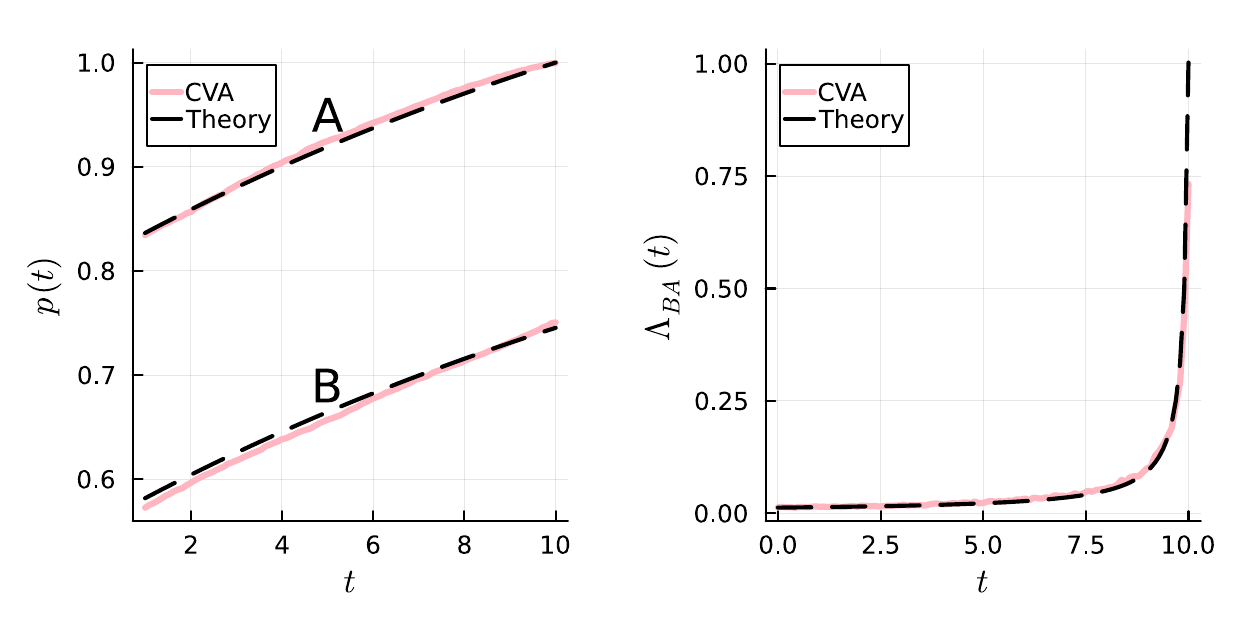}\caption{\label{fig:Intro:effective_interaction-1}. Comparison between the
exact calculations (dashed black lines) and results obtained using the \Causalityacronym{} (red lines) for a SI model ($\lambda=0.05$ and $\gamma=1/2$) with two individuals A and B, where A is observed infected at the final time $T=10$.
\textit{Left}: Marginal probabilities of being infected for individuals A  and B as a function of time. Due to the observation on A at time $T$, the two marginals are different: in particular, individual A's marginal reaches 1 at time $T$ in such a way as to be consistent with the observation. \textit{Right}: Effective
infection rate $\Lambda_{\text{B}\text{A}}(t)$ as function of time. The divergence is due to the observation on individual A at time $T$. }
\end{figure*}

The exact posterior distribution \eqref{eq:Intro:N=00003D2posterior} can be expressed in the form of a generalized SI model with asymmetric probabilities $\Gamma_{\text{A}}$ and $\Gamma_{\text{B}}$ that the individuals are already infected at the initial time and time-dependent infection rates $\Lambda_{\text{A}\text{B}}(t)$ and $\Lambda_{\text{B}\text{A}}(t)$. A schematic representation of the role played by these parameters as compared with the ones of the original SI model is provided in Fig. \ref{fig:SIn=2} (\textit{right}). Such a generalized SI model can be used as an ansatz distribution $Q_\theta$ in the \Causalityacronym{}. The minimum of the corresponding free energy is obtained when the generalized parameters $\Gamma_{\text{A}},\Gamma_{\text{B}}, \Lambda_{\text{A}\text{B}}(t)$ and $\Lambda_{\text{B}\text{A}}(t)$ assume exactly the form derived above in Eqs. \eqref{eq:Intro:N=00003D2posterior_seedA},\eqref{eq:Intro:N=00003D2posterior_seedB},\eqref{eq:Intro:N=00003D2postInteractionAB} and \eqref{eq:Intro:N=00003D2postInteractionBA}. It follows that the \Causalityacronym{} provides a formally exact solution to the inference problem under study. 
This is shown in Figure \ref{fig:Intro:effective_interaction-1}, where the exact solution of the model is compared with the solution found using the \Causalityacronym{}.

\section{Probabilistic description of the SEIR model with observations}\label{sec:appendix_SEIR}
The Susceptible-Exposed-Infected-Recovered (SEIR) model is a generalization of the SI model in which incubation ($E$) and recovery ($R$) are included. The only allowed transitions are $S\to E$, $E\to I$, $I\to R$. 
A transmission event by an infected individual $j$ brings a susceptible individual $i$ into the $E$ state with a rate $\lambda_{ji}$, then the transition to $I$ occurs independently of the rest of the system with rate $\nu_{i}$. Finally, an infected individual $i$ recovers, again independently of the others, with rate $\mu_{i}$. In this model, the trajectory of the
epidemic process is fully specified by three times: $t_{i}^{E}\leq t_{i}^{I}\leq t_{i}^{R}\in\mathbb{R}_{\geq 0}$, representing the times in which individual $i$ enters the states $E$, $I$ and $R$, respectively. For an initially infected individual 
  $t_{i}^{E}=t_{i}^{I}=0$. In this notation, the probability weight of the
trajectory $\boldsymbol{t}=\left\{ (t_{i}^{E},t_{i}^{I},t_{i}^{R})\right\} _{i=1}^{N}$
can be written as

\begin{equation}
\mathbb{P}\left[\boldsymbol{t}\right] =\prod_{i}\Bigg[\gamma\delta\left(t_{i}^{E}\right)\delta\left(t_{i}^{I}\right)+\left(1-\gamma\right)\Lambda\left(\sum_{j\neq i}\left[t_{j}^{I}\leq t\leq t_{j}^{R}\right]\lambda_{ji}\left(t\right),t_{i}^{E}\right) \Lambda\left(\nu_{i}\mathbb{I}\left[t_{i}^{E}\leq t\right],t_{i}^{I}\right)\Bigg]\Lambda\left(\mu_{i}\mathbb{I}\left[t_{i}^{I}\leq t\right],t_{i}^{R}\right)
\end{equation}

Test-based observations can be defined by generalizing the argument discussed in Section \ref{sec:EpidemicRiskAss} in the main text for the SI model. Once again, the simplest realization of test $r$ consists of a noisy evaluation of the individual's infection state, whose outcome obeys the following stochastic expression, conditioned w.r.t. the time trajectory of individual $i$:
\begin{subequations}
\begin{align}
\nonumber\mathbb{P}\left[r=+|{t}_{i}\right] & =\left(1-p_{\rm{FNR}}\right)\,\mathbb{I}\left[t_{i}^{E}\leq t < t_{i}^{R} \right]+p_{\rm{FPR}}\left( \mathbb{I}\left[t< t_{i}^{E}\right] + \mathbb{I}\left[t\geq t_{i}^{R}\right] \right)\\
\mathbb{P}\left[r=-|{t}_{i}\right] & =p_{\rm{FNR}}\,\mathbb{I}\left[t_{i}^{E}\leq t < t_{i}^{R} \right]+\left(1-p_{\rm{FPR}}\right)\,\left( \mathbb{I}\left[t< t_{i}^{E}\right] + \mathbb{I}\left[t\geq t_{i}^{R}\right] \right)
\end{align}
\end{subequations}
where $p_{\rm{FPR}}$ (resp. $p_{\rm{FNR}}$) is the false positive (resp. negative) ratio of the test. Here it was implicitly assumed that the test outcome does not depend on any acquired immunity.

Finally, the definition of the risk measure in Eq. \eqref{eq:riskSI} in the main text can be straightforwardly generalized to the SEIR model 
\begin{equation}
\mathbb{P}\left[x_{i}\left(t\right)=I|\mathcal{O}\right]=\int d \boldsymbol{t}\mathbb{I}\left[t_{i}^I\leq t <t_{i}^{R}\right]\mathbb{P}\left[\boldsymbol{t}|\mathcal{O}\right] \label{eq:riskSEIR}
\end{equation}
where $\int d \boldsymbol{t}$ denotes the integral over all transition
time triplets $(t_{1}^{E},t_{1}^{I},t_{1}^{R}),\dots,(t_{N}^{E},t_{N}^{I},t_{N}^{R})$.

\section{Parametrization in the \label{sec:Appendix:Parametrizations} \Causalityname{}}

A strong advantage of \Causalityacronym{} is its flexibility in the parametrization
of the solution ansatz. It is sufficient, indeed, to weakly generalize the prior model in order to obtain the form of the $Q_{\theta}(\boldsymbol{x})$
to optimize. This section describes the parametrizations used for the three models whose results are presented in the main text, namely the Random Walk and the epidemic SI and SEIR models.

\paragraph*{Random Walk --}

 The Random Walk generative model is a discrete-time stochastic process, where at each time the walker chooses to jump right or left with a uniform probability $1/2$. As discussed in Section \ref{sec:RW_maintext} of the main text, the goal is to characterize the probability distribution of a constrained random walk. The \Causalityacronym{} simply consists in assuming a more comprehensive parametrization of the generative model, in a way that, the conditioned distribution thus obtained still describes a Random Walk. In particular, instead of fixing the jump probability to $1/2$, it is allowed to take different, heterogeneous values, depending on time and space. The parameters of the $Q_{\theta}(\boldsymbol{x})$ are then simply identified with probabilities $r_{i}^{t}$ to jump to the right for a walker that at
time $t$ is in position $x(t)=i$. The ansatz has therefore the following form,
\begin{equation}
Q_{\theta}(\boldsymbol{x})=\prod_{t=0}^{T-1}\left[r_{x(t)}^{t}\delta_{x\left(t+1\right),x(t)+1}+\left(1-r_{x(t)}^{t}\right)\delta_{x\left(t+1\right),x(t)-1}\right]    
\end{equation}
which describes an inhomogeneous time-dependent random walk. \Causalityacronym{} reduces to optimize $Q_{\theta}$ over the probabilities $\{r_{i}^{t}\}_{i=1,\dots,N}^{t=1,\dots,T}$. 
\paragraph*{Homogeneous Markovian SI model --}
To characterize a homogeneous and Markovian SI dynamic one needs to specify the constant infection rate $\lambda$, together with the probability $\gamma$ of being the sources of the infection at the initial time. Thus, the probability of the infection times in Eq. \eqref{eq:Intro:SI_continuous-2} of the main text reads
\begin{align}
\mathbb{P}\left[\boldsymbol{t}\right] & = \prod_{i}\left\{ \gamma\delta\left(t_{i}\right)+\left(1-\gamma\right)\Lambda\left(\sum_{j\neq i}\mathbb{I}\left[t_{j}\leq t\right]\lambda,t_{i}\right)\right\}.\label{eq:appendix:homogMarkovSI}
\end{align}
The Causality ansatz for this model is obtained by introducing, for each individual $i$, effective infection rates $\{\lambda_{i}(t)\}_{i=1,\dots,N}$ and zero-patient probabilities $\{\gamma_{i}\}_{i=1,\dots,N}$ respectively.
In other words, the conditioned, and constrained, SI dynamical model is approximated with an unconditioned but inhomogeneous and time-dependent SI model. 
Each $\lambda_i(t)$ refers to the `incoming` infection rate at site $i$ at time $t$, namely, we impose that $\lambda_{ji}\left(t\right) = \lambda_{i}\left(t\right)$ for every $j\in\partial i$.  
The benefit carried by this parametrization is twofold: (i) this individual infection allows us to simplify the calculations and (ii) it guarantees `outgoing` heterogeneous infection rates. A simple example can be used to illustrate this parametrization. Suppose to consider an individual $i$ that has only two contacts with $j_1$ and $j_2$, observed to be respectively $I$ and $S$ at time $\tau$. Since $j_2$ is susceptible at time $\tau$, none of its previous contacts has infected it at previous times; in our parametrization, this case can be encoded by setting $\lambda_{j_{2}}\left(t\right) = 0$ for $t < \tau$. To cope with the observation of the state of $j_{1}$ one can tune  $\lambda_{j_1}(t)$ (for $t < \tau$) to sufficiently high values to guarantee that $j_{1}$ is in state $I$ at the observation time. 
In the present work, self-infection rates $\{\omega_{i}\}_{i=1,\dots,N}$ are also introduced, that is the rates with which individuals can get infected without any contact with an infectious individual. Although this transition is not contemplated in the generative model, it is conveniently included to satisfy some constraints associated with the observation of infected individuals, that are hard to justify through the infection rates alone. 
In this case, the \Causalityacronym{} ansatz reads
\begin{equation}
\label{eq:app:param:SIansatz}
Q_\theta(\boldsymbol{x}) = \prod_{i} \Bigg\{ \gamma_{i}\delta\left(t_{i}\right)+\left(1-\gamma_{i}\right) \Lambda\left(\sum_{j\in\partial i}\mathbb{I}\left[t_{j}^{I}\leq t\right]\lambda_{i}\left(t\right) + \omega_i\left(t\right),t_{i}\right)\Bigg\},
\end{equation}
Notice that Eq. \eqref{eq:app:param:SIansatz} is almost identical to \eqref{eq:Intro:SI_continuous-2} of the main text, except for the presence of the self-infection rates.
The SI model considered here is continuous in time, so that the variational parameters
$\lambda_{i}$ and $\omega_i$ should in principle be treated as continuous functions over time as well. 

Since it is not possible to optimize over a function defined on a finite real domain, it is more convenient to express such rates using a family of functions, defined by a set of parameters, and optimize over the latter. A simple choice - adopted in all the results presented in the main text - is a Gaussian-like (not normalized) function for each site-dependent rate $\lambda_i$ and $\omega_i$. Each Gaussian rate will thus depend on three parameters, namely the value of the peak (labeled using $\mu$), the standard deviation ($\sigma$), and a scale factor ($p$). For instance, the infection rate $\lambda_{i}(t)$ reads
\begin{equation}
\lambda_{i}\left(t\right)=\lambda_{i}^{p}\exp\left[-\left(\frac{t-\lambda_{i}^{\mu}}{\lambda_{i}^{\sigma}}\right)^{2}\right], \label{eq:GaussianRate}
\end{equation}
and analogous expressions hold for the self-infection rates $\omega_i (t)$ with parameters $\left( \omega_{i}^{p},\omega_{i}^{\mu},\omega_{i}^{\sigma} \right)$. With this choice, the total set of parameters to optimize over is given by:
\begin{equation}
\theta=\{\lambda_{i}^{p},\lambda_{i}^{\mu},\lambda_{i}^{\sigma},\omega_{i}^{p},\omega_{i}^{\mu},\omega_{i}^{\sigma},\gamma_{i}\}_{i=1,\dots,N}    
\end{equation}

\paragraph*{Homogeneous Non-Markovian SI model --}

A more realistic scenario is that depicted by the homogeneous Non-Markovian SI model, where the infection rates are not individual-dependent (e.g. spatially homogeneous) but they can vary in time. In this case, Eq. \eqref{eq:Intro:SI_continuous-2} of the main text slightly modifies as
\begin{align}
\mathbb{P}\left[\boldsymbol{t}\right] & = \prod_{i}\left\{ \gamma\delta\left(t_{i}\right)+\left(1-\gamma\right)\Lambda\left(\sum_{j\neq i}\mathbb{I}\left[t_{j}\leq t\right]\lambda(t-t_j),t_{i}\right)\right\},\label{eq:appendix:homogNon-MarkovSI}
\end{align}
This is quite a realistic hypothesis.
In fact, when an individual gets infected, its infectivity is not constant
in time in real situations but it depends on the viral load of the infectious individuals. Typically, infectivity is very low in a first time-window after the contagion, then increases according to time scales that depend on the type of pathogens, and finally, it decreases to zero. In particular, the generative rate $\lambda$ depends
on the time elapsed since infection, i.e. $\lambda(t-t_{j})$. This introduces a Non-Markovian character to the dynamic because in order to
know the configuration at time $t+dt$ it is not sufficient to know the state at time $t$, but a memory of all the infection times of each individual must be kept. Even though the generative model becomes
more complex with respect to the Markovian case described above, the inference with \Causalityacronym{} has almost no modifications. For this reason, the same inferential parameters introduced in the Markovian case are used, with the same interpretation. The only difference is that the effective interaction rate is now defined as \begin{equation}
\lambda_{ji}^{eff}(t|t_{j}):=\frac{\lambda(t-t_{j})\lambda_{i}(t)}{\lambda_{0}}\label{eq:lambda_eff_nonmarkovSI}
\end{equation}
(where $\lambda_{0}$ is a rate needed to preserve the correct dimension, but it can be numerically set to 1). It is worth noting that the effective infection rate encompasses both the time-dependent infectivity of individual $j$ (which is given by the generative model) and the susceptibility of $i$ to be infected (which instead is learned by \Causalityacronym{}).
The formula for the \Causalityacronym{} ansatz is therefore
\begin{equation}
Q_\theta(\boldsymbol{x}) = \prod_{i} \Bigg\{ \gamma_{i}\delta\left(t_{i}\right)+\left(1-\gamma_{i}\right) \times\Lambda\left(\sum_{j\in\partial i}\mathbb{I}\left[t_{j}^{I}\leq t\right]\lambda^{eff}_{ji}\left(t|t_j\right) + \omega_i\left(t\right),t_{i}\right)\Bigg\},\label{eq:app:param:Non-MarkovSIansatz}
\end{equation}.

\paragraph*{SEIR model --}

To cope with the SEIR model, it suffices to slightly modify the formalism introduced for the SI models described above. Together with the infection rates and the source probabilities, it is necessary to include, in the generative model, the latency rate $\nu$ and the recovery rate $\mu$, associated with the transitions $E\to I$ and $I\to R$, respectively. Hence, the parameters $\theta$ now encompass, for each individual $i$, the zero-patient probability, an infection rate function $\lambda_{i}(t)$, an auto-infection rate $\omega_{i}(t)$, and the recovery and latency rates $\mu_{i}(t)$ and $\nu_{i}(t)$. In the results shown in Section \ref{sec:EpidemicRiskAss} of the main text, all these rates are parametrized using Gaussian functions, as in Eq. \eqref{eq:GaussianRate}. \\
As a final remark, it is worth noting that, for both the SI and the SEIR models, the total number of parameters
used by the \Causalityacronym{} scales with $N$ (in particular, their number is $7N$ for the SI and
$13N$ for the SEIR).

\section{Sampling in the \Causalityname{} \label{sec:Appendix:Sampling}}

Sampling from the posterior distribution is a difficult task in general.
The \Causalityacronym{} allows one to approximate the posterior with a distribution
$Q_{\theta}$ from which, instead, it is possible to sample efficiently. The
sampling process is crucial for performing the gradient descent described
in Appendix \ref{sec:Appendix:Descent}. The present  Appendix provides all the implementation details required to sample the probabilistic models considered in the paper. 

\paragraph*{Random walk --}

In Appendix \ref{sec:Appendix:Parametrizations}, 
the ansatz $Q$ takes the form of an inhomogeneous time-dependent random walk. suppose a trajectory  $x=(x^{0},x^{1},\dots,x^{T})$ has to be drawn from $Q$, where
$x^{t}$ is the position of the walker at time $t$ and $x^{0}=0$.
To sample $x$, it is sufficient to start from state $x^0$ and repeat, up to the final time $T$, the temporal update $x^{t+1}=x^{t}+\Delta x$, where $\Delta x\in\{-1,+1\}$
is a binary random variable. The probability $r_{x^{t}}^{t}$, that
$\Delta x =1$, is the probability for a walker to jump to the right
at time $t$ starting from position $x^{t}$. 

\paragraph*{SI model --}

The parametrization used to infer the conditioned SI model is an SI model itself with self-infection probability, as described in Appendix \ref{sec:Appendix:Parametrizations}. A Gillespie algorithm, described in detail in Algorithm \ref{alg:Sampling}, is used to sample from the SI model distribution. The idea behind the algorithm is to store in a queue $L=(L_{1},L_{2},\dots,L_{N})$
the infection times $L_i$ of all the individuals. To this end, the first step consists of sampling all the zero-patient(s), i.e. all individuals $i$ such that $L_{i}=0$.
Secondly, self-infection events are sampled for each individual $i$ which is not a zero patient, 
with a random variable $t_{i}$ extracted from the self-infection
distribution $\omega_{i}(t)$. 

Finally, contagion events are sampled by extracting from the queue the individual having the minimum value of $L$.
If an individual $i$ tries to infect another individual $j$ at time
$t_{ij}$, then $j$ updates its infection time by taking the minimum
between $t_{ij}$ and its current value of $L_{j}$. In this way, the list
is updated recursively. \\
As a final remark, notice that to ensure a proper epidemic process to take place, at least one zero patient is needed. Therefore, it is necessary to sample events with $L_i=0$ by constraining a positive number $n_{zp}$ of zero-patients. This procedure can be easily carried out recursively. For instance, starting from node $1$, we set it to be a patient-zero with a probability
\begin{align}
\mathbb{P}(L_{1}=0|n_{zp}>0) =\frac{\mathbb{P}(L_{1}=0,n_{zp}>0)}{\mathbb{P}\left(n_{zp}>0\right)}=\frac{\mathbb{P}(L_{1}=0)}{\mathbb{P}\left(n_{zp}>0\right)} =\frac{\gamma_{1}}{1-\prod_{i=1}^{N}\left(1-\gamma_{i} \right)}
\end{align}
Then, the next node (say number $2$) is set to be infected at $t=0$ with a probability depending on $L_1$, namely:
\begin{equation}
\mathbb{P}(L_{2}=0|L_{1},n_{zp}>0) =\begin{cases}
\gamma_{2}  &\text{if \ensuremath{L_{1}=0}}\\
\gamma_{2}/\left(1-\prod_{i\geq2}(1-\gamma_{i})\right) & \text{if }L_{1}>0
\end{cases}\label{eq:sampling_node2_Li0}
\end{equation}and the above strategy is iterated for every $L_{k}$. Intuitively, from Eq. \eqref{eq:sampling_node2_Li0} it follows that as soon as one zero patient is sampled, the remaining individuals' state is extracted independently with probability $\gamma_i$.
\begin{algorithm}
\begin{itemize}
\item Initialize a queue $L$ 
\item \textbf{Loop }over population: $i:1,\dots,N$
\begin{itemize}
\item $L[i]\assignmentop 0$ with probability: $\gamma_{i}/\left(1 -\prod_{j\geq i}(1-\gamma_{j})\right)$
\item \textbf{if $L[i]=0$ break the loop}
\end{itemize}
\item \textbf{Loop }over the remaining population:
\begin{itemize}
\item $L[i] \assignmentop 0$ with probability $\gamma_{i}$
\end{itemize}
\item \textbf{Loop }over the non zero-patients (variable of the loop: $i$)
\begin{itemize}
\item $L[i]\assignmentop t_{i}$ where $t_{i}$ is extracted from the self-infection rate $\omega_{i}$
\end{itemize}
\item \textbf{Loop }over the queue $L$ entering by infection time (from the smallest to the highest)
\begin{itemize}
\item Call $i$ the element extracted from $L$
\item \textbf{Loop }over $j\in\partial i$
\begin{itemize}
\item Extract the time $t_{ij}$ at which $i$ tries to infect $j$ by extracting from the Gaussian distribution of $\lambda_{ij}^{eff}$ 
\item $L[j]\assignmentop \min\{L[j],t_{ij}\}$
\end{itemize}
\item Throw away $i$ from the queue $L$ and save $L[i]$ as the infection
time of $i$.
\end{itemize}
\item \textbf{Return }the list of infection times.
\end{itemize}
\caption{Sampling the effective (Non-Markovian) SI model \label{alg:Sampling} }
\end{algorithm}

\paragraph*{SEIR model --}

For the SEIR model, the sampling procedure is a straightforward generalization of the one just discussed. The only difference is that the queue contains all the transition time triplets $t_i^{E},t_i^{I},t_i^{R}$ for each individual.
{\textbf{Remark.} Notice that, since each sample is independent to the other, sampling can be performed in parallel. This implies, as shown in the next section, that CVA itself is a parallelizable algorithm.}
\section{The \Causalityname{}: Gradient Descent\label{sec:Appendix:Descent}}

In the \Causalityname{}, the parameters $\theta$ of the approximation are determined by minimizing the following KL divergence:
\begin{equation}
D_{KL}(Q_{\theta}||\mathbb{P}\left[\boldsymbol{x} | \mathcal{O} \right])=\int d\boldsymbol{x}\, Q_{\theta}\left(\boldsymbol{x}\right) \log \left(\frac{Q_{\theta}(\boldsymbol{x})}{\mathbb{P}\left[\boldsymbol{x} | \mathcal{O}\right]} \right),    
\end{equation}
where $Q_{\theta}$ denotes the Causality ansatz. The integration is formal. If $\mathbb{P}\left[ \boldsymbol{x}| \mathcal{O}\right]$ is continuous,
then it corresponds to a Lebesgue integral, otherwise, it is a sum over the discrete state $\boldsymbol{x}$. 

\paragraph*{Gradient Descent --}

The KL divergence can be minimized by performing a gradient descent in the $\theta$-parameter space. The parameters at iteration $k+1$ read
\begin{equation}
\theta^{(k+1)}=\theta^{(k)}-\epsilon\nabla_{\theta}D_{KL}(Q_{\theta}||\mathbb{P\left[\boldsymbol{x} | \mathcal{O}\right]}) \label{eq:grad}
\end{equation}
where $\epsilon$ is the learning rate. Before entering the calculation
of the gradient, however, let
us observe that, in general, $\theta$ contains parameters which
have different scales, and, therefore, a straightforward implementation of eq. \eqref{eq:grad} may be inefficient. To solve this issue, we resort to a sign descender method \cite{bernstein_signsgd_2018}, a simple scale-free descender. Briefly, it consists in moving in the direction of the
partial derivative with respect to each parameter without using the information of the gradient's magnitude. 

Finally, the update rule for a parameter $\theta_i$ used in this work has the following expression:
\begin{equation}
\theta_{i}^{(k+1)}=\theta_{i}^{(k)}\left[1-\varepsilon\text{\,sign}\left(\partial_{\theta_{i}}D_{KL}(Q_{\theta}||\mathbb{P\left[\boldsymbol{x} | \mathcal{O}\right]})\right)\right]\label{eq:Appendix:updateSignDesc}
\end{equation}
It is now necessary to evaluate the derivatives of  $D_{KL}(Q_{\theta}||\mathbb{P}\left[\boldsymbol{x} | \mathcal{O} \right])$,
\begin{align}
\partial_{\theta_{i}}D_{KL}(Q_{\theta}||\mathbb{P\left[\boldsymbol{x} | \mathcal{O}\right]}) & =\partial_{\theta_{i}}\int d\boldsymbol{x}Q_{\theta}(x)\log\frac{Q_{\theta}(\boldsymbol{x})}{\mathbb{P\left[\boldsymbol{x} | \mathcal{O}\right]}}=\nonumber \\
 & =\partial_{\theta_{i}}\int d\boldsymbol{x}Q_{\theta}(\boldsymbol{x})\log\frac{Q_{\theta}(\boldsymbol{x})\mathbb{P}[\mathcal{O}]}{\mathbb{P}[\boldsymbol{x}]\mathbb{P}\left[\mathcal{O}|\boldsymbol{x}\right]}=\nonumber \\
 & =\int d\boldsymbol{x}\partial_{\theta_{i}}\left(Q_{\theta}(\boldsymbol{x})\log\frac{Q_{\theta}(\boldsymbol{x})}{\mathbb{P}\left[\boldsymbol{x}\right]\mathbb{P}\left[\mathcal{O}|\boldsymbol{x}\right]}\right)+\log \mathbb{P}\left[\mathcal{O}\right]\partial_{\theta_{i}}\int d\boldsymbol{x}\left(Q_{\theta}(\boldsymbol{x})\right)=
 \nonumber \\
 & =\int d \boldsymbol{x}\partial_{\theta_{i}}\left(Q_{\theta}(\boldsymbol{x})\log\frac{Q_{\theta}(\boldsymbol{x})}{\mathbb{P}\left[\boldsymbol{x}\right] \mathbb{P}\left[\mathcal{O}|\boldsymbol{x}\right]}\right)\label{eq:Appendix:descent:eliminatingPOfrom_gradient}
\end{align}
in which the dependency on $P\left[\mathcal{O}\right]$ was neglected.
We observe that 
\begin{equation}
\int d\boldsymbol{x}\,\partial_{\theta_{i}}\left(Q_{\theta}(\boldsymbol{x})\log Q_{\theta}(\boldsymbol{x})\right) =\int d\boldsymbol{x}\,\partial_{\theta_{i}}Q_{\theta}(\boldsymbol{x})\log Q_{\theta}(\boldsymbol{x})+ \int d\boldsymbol{x}\,Q_{\theta}(\boldsymbol{x})\partial_{\theta_{i}}\log Q_{\theta}(\boldsymbol{x})
\end{equation}
In the above expression, the second term of the r.h.s. is $0$, since
\begin{equation}
\int d\boldsymbol{x}Q_{\theta}(\boldsymbol{x})\partial_{\theta_{i}}\log Q_{\theta}(\boldsymbol{x})=\int d\boldsymbol{x}\partial_{\theta_{i}}Q_{\theta}(\boldsymbol{x})= \partial_{\theta_{i}}\int d \boldsymbol{x}Q_{\theta}(\boldsymbol{x})=0,
\end{equation}
therefore
\begin{align}
\int d\boldsymbol{x}\partial_{\theta_{i}}\left(Q_{\theta}(\boldsymbol{x})\log Q_{\theta}(\boldsymbol{x})\right)  &=\int d\boldsymbol{x}\left(\partial_{\theta_{i}}Q_{\theta}(\boldsymbol{x})\right)\log Q_{\theta}(\boldsymbol{x})\nonumber \\
 & =\int d\boldsymbol{x}\frac{Q_{\theta}(\boldsymbol{x})}{Q_{\theta}(\boldsymbol{x})}\left(\partial_{\theta_{i}}Q_{\theta}(\boldsymbol{x})\right)\log Q_{\theta}(\boldsymbol{x})=\nonumber \\
 & =\int d\boldsymbol{x}\,Q_{\theta}(\boldsymbol{x})\left(\partial_{\theta_{i}}\log Q_{\theta}(\boldsymbol{x})\right)\log Q_{\theta}(\boldsymbol{x})=\nonumber \\
 & =\left\langle \log Q_{\theta}(\boldsymbol{x})\partial_{\theta_{i}}\log Q_{\theta}(\boldsymbol{x})\right\rangle _{Q_{\theta}}\label{eq:Appendix:QlogQ_trick}
\end{align}
In conclusion, the derivative of the $KL$ divergence reads
\begin{equation}
\partial_{\theta_{i}}D_{KL}(Q_{\theta}||\mathbb{P}\left[ \boldsymbol{x} | \mathcal{O}\right])=\left\langle \log\frac{Q_{\theta}(\boldsymbol{x})}{\mathbb{P}\left[\boldsymbol{x}\right]\mathbb{P}\left[\mathcal{O}|\boldsymbol{x}\right]}\partial_{\theta_{i}}\log Q_{\theta}(\boldsymbol{x})\right\rangle _{Q_{\theta}}\label{eq:Appendix:derivativeKLparams}
\end{equation}
The problem of calculating the derivative of the KL divergence
has been reduced to calculate the derivative of the logarithm of the
ansatz and then to average over the ansatz. In principle, one may need to calculate
\begin{equation}
\partial_{\theta_{i}}\log Q_{\theta}(\boldsymbol{x})=\sum_{j=1}^{N}\partial_{\theta_{i}}\log q_{j}(x_{j},x_{\partial j})
\end{equation}
but it is reasonable to assume that the dependency of each parameter $\theta_{i}$ occurs only on the term $\log q_{i}$. This hypothesis is true for all the examples and applications presented in this work, as described in the previous Appendix.
Therefore
\begin{equation}
\partial_{\theta_{i}}D_{KL}(Q_{\theta}||\mathbb{P} \left[\boldsymbol{x} | \mathcal{O} \right])= \left\langle \log\frac{Q_{\theta}(\boldsymbol{x})}{\mathbb{P}\left[\boldsymbol{x}\right]\mathbb{P}\left[\mathcal{O}|\boldsymbol{x}\right]}\partial_{\theta_{i}}\log q_{i}(x_{i},x_{\partial i})\right\rangle _{Q_{\theta}}\label{eq:Appendix:derivative_of_KLbeforeVR}
\end{equation}
A last manipulation consists in subtracting the quantity $0= \left\langle \log\frac{Q_{\theta}(\boldsymbol{x})}{\mathbb{P}\left[\boldsymbol{x}\right]\mathbb{P}\left[\mathcal{O}|\boldsymbol{x}\right]}\right\rangle _{Q_{\theta}} \left\langle \partial_{\theta_{i}}\log Q_{\theta}(\boldsymbol{x})\right\rangle _{Q_{\theta}}$  to equation \eqref{eq:Appendix:derivative_of_KLbeforeVR} in order to obtain:

\begin{equation}
\partial_{\theta_{i}}D_{KL}(Q_{\theta}||\mathbb{P} \left[\boldsymbol{x} | \mathcal{O} \right])= \left\langle\left( \log\frac{Q_{\theta}(\boldsymbol{x})}{\mathbb{P}\left[\boldsymbol{x}\right]\mathbb{P}\left[\mathcal{O}|\boldsymbol{x}\right]}-\left\langle \log\frac{Q_{\theta}(\boldsymbol{x})}{\mathbb{P}\left[\boldsymbol{x}\right]\mathbb{P}\left[\mathcal{O}|\boldsymbol{x}\right]}\right\rangle _{Q_{\theta}} \right)\partial_{\theta_{i}}\log q_{i}(x_{i},x_{\partial i})\right\rangle _{Q_{\theta}}.\label{eq:Appendix:derivative_of_KL}
\end{equation}
This manipulation is called \textit{variance reduction} \cite{StatMech_with_Autoreg} and is known to facilitate the gradient descent.
Averages are evaluated by sampling from $Q_{\theta}$ using the implementation details given in Appendix \ref{sec:Appendix:Sampling}.  {Since sampling can be performed in parallel, also the gradient computation is parallelizable, due to the form of equation \eqref{eq:Appendix:derivative_of_KL}.} The learning process
is therefore the following:
\begin{enumerate}
\item The parameters are initialized to $\theta^{0}$ (we initialize them
with the values of the generative model, where known. For example, in the random walk example
the jump probabilities are all initialized to $1/2$).
\item The derivatives of the KL divergence are performed by sampling from
$Q_{\theta_{0}}$ and through Eq. \eqref{eq:Appendix:derivative_of_KL}.
\item The parameters are updated to $\theta^{1}$ by using \eqref{eq:Appendix:updateSignDesc}.
\item Steps 2. and 3. are repeated.
\item The process stops when $D_{KL}(Q_{\theta}||\mathbb{P}\left[\boldsymbol{x} | \mathcal{O} \right])$ does not significantly change after two consecutive iterations.
\end{enumerate}
In the evaluation of the KL divergence (or its derivatives), it is necessary to evaluate quantities that include $\log \mathbb{P}\left(\mathcal{O} \mid \boldsymbol{x}\right)$. When $\mathcal{O}$ imposes a hard constraint on the trajectory - as it happens in the epidemic models with noiseless observations - the violation of one of the constraints (which is likely to occur especially in the first stages of the gradient descent) would result in $\log \mathbb{P}\left(\mathcal{O}\mid \boldsymbol{x}\right) \to -\infty $. To avoid this issue, it is convenient to relax (soften) all the constraints: a small acceptance $p\sim10^{-10}$ is introduced, in such a way that $\log \mathbb{P}\left(\mathcal{O}\mid \boldsymbol{x}\right)$ has always non-diverging values. As a consequence, the distribution
$Q_{\theta}(\boldsymbol{x})$ resulting from the KL minimization gives a non-zero (yet very small) probability to events $\boldsymbol{x}$
which do not satisfy the constraint.

\section{Inference of Hyperparameters \label{sec:hyperparams_appendix}}
As explained in Section \ref{sec:EpidemicRiskAss} in the main text, the term \emph{hyperparameters} refers to parameters of a generative model. For example,
the Random Walk presented in Section \ref{sec:RW_maintext} in the main text has a unique hyperparameter, namely the probability of a right jump, which is set to $1/2$ for all times and sites. In the SI model, the hyperparameters
are the zero-patient probability $\gamma$ and the infection rate
$\lambda$ while in the SEIR model, we need to add to the hyperparameters set the latency
rate $\nu$ and the recovery rate $\mu$. Usually, in inference problems, they are not known as the only source of information, in fact, are the observations. However, inferring the hyperparameters from observations is crucial to ensure an effective solution to the underlying inference problem. The \Causalityacronym{} allows one to estimate them. Let us call the set of hyperparameters $\theta^{p}$. First, notice that in general, not
only the generating distribution $\mathbb{P}\left[ \boldsymbol
{x} \right]$, but also the ansatz $Q_{\theta,\theta^{p}}$ can depend on the hyperparameters $\theta^{p}$. This dependency might be present when using part of the generative model in the ansatz distribution. The example implemented in this paper is the case of the Non-Markovian SI model (see Appendix \ref{sec:Appendix:Parametrizations}). 
In that case, the infectivity function depends on both the inferred
parameters ($\{\lambda_{i}\}\in\theta$) and the relative infectivity
of the generative model ($\lambda\in\theta^{p}$), as shown in Eq. \eqref{eq:lambda_eff_nonmarkovSI}.
To optimize the parameters, we will use a Maximum Likelihood (ML) approax, maximizing $\log\mathbb{P}\left[O|\theta^{p}\right]$ with respect to $\theta^{p}$:
\begin{align*}
\arg\max_{\theta^{p}}\log\mathbb{P}\left[O|\theta^{p}\right] & \approx\arg\min_{\theta^{p}}\min_{\theta}\left[D_{KL}\left(Q_{\theta,\theta^{p}}||\mathbb{P}\left[\boldsymbol{x}|O,\theta^{p}\right]\right)-\log\mathbb{P}\left[O|\theta^{p}\right]\right]\\
 & =\arg\min_{\theta^{p}}\min_{\theta}\int d\boldsymbol{x}Q_{\theta,\theta^{p}}\left[\boldsymbol{x}\right]\log\frac{Q_{\theta,\theta^{p}}\left[\boldsymbol{x}\right]}{\mathbb{P}\left[O|\boldsymbol{x},\theta^{p}\right]\mathbb{P}\left[\boldsymbol{x}|\theta^{p}\right]}
\end{align*}

Where the $\approx$ becomes an equality if the Ansatz $Q_{\theta,\theta^{p}}$
is general enough to represent $\mathbb{P}\left[\boldsymbol{x}|O,\theta^{p}\right]$.
So our task is to minimize $\int d\boldsymbol{x}Q_{\theta,\theta^{p}}\left[\boldsymbol{x}\right]\log\frac{Q_{\theta,\theta^{p}}\left[\boldsymbol{x}\right]}{\mathbb{P}\left[O|\boldsymbol{x},\theta^{p}\right]\mathbb{P}\left[\boldsymbol{x}|\theta^{p}\right]}$
with respect to both $\theta$ and $\theta^{p}$. In doing so, we
will simultaneously find our best approximation $Q_{\theta,\theta^{p}}$
of the posterior $\mathbb{P}\left[\boldsymbol{x}|O,\theta^{p}\right]$
. Let us compute gradients:

\begin{align}
\nabla_{\theta}\int d\boldsymbol{x}Q_{\theta,\theta^{p}}\left[\boldsymbol{x}\right]\log\frac{Q_{\theta,\theta^{p}}\left[\boldsymbol{x}\right]}{\mathbb{P}\left[O|\boldsymbol{x},\theta^{p}\right]\mathbb{P}\left[\boldsymbol{x}|\theta^{p}\right]}= & \nabla_{\theta}\int d\boldsymbol{x}Q_{\theta,\theta^{p}}\left[\boldsymbol{x}\right]\log\frac{Q_{\theta,\theta^{p}}\left[\boldsymbol{x}\right]}{\mathbb{P}\left[O|\boldsymbol{x},\theta^{p}\right]\mathbb{P}\left[\boldsymbol{x}|\theta^{p}\right]}\\
= & \int d\boldsymbol{x}\nabla_{\theta}Q_{\theta,\theta^{p}}\left[\boldsymbol{x}\right]\log\frac{Q_{\theta,\theta^{p}}\left[\boldsymbol{x}\right]}{\mathbb{P}\left[O|\boldsymbol{x},\theta^{p}\right]\mathbb{P}\left[\boldsymbol{x}|\theta^{p}\right]}+\nonumber \\
 & +\int d\boldsymbol{x}Q_{\theta,\theta^{p}}\left[\boldsymbol{x}\right]\nabla_{\theta}\log\frac{Q_{\theta,\theta^{p}}\left[\boldsymbol{x}\right]}{\mathbb{P}\left[O|\boldsymbol{x},\theta^{p}\right]\mathbb{P}\left[\boldsymbol{x}|\theta^{p}\right]}\label{eq:cancelled}\\
= & \int d\boldsymbol{x}Q_{\theta,\theta^{p}}\left[\boldsymbol{x}\right]\nabla_{\theta}\log Q_{\theta,\theta^{p}}\left[\boldsymbol{x}\right]\log\frac{Q_{\theta,\theta^{p}}\left[\boldsymbol{x}\right]}{\mathbb{P}\left[O|\boldsymbol{x},\theta^{p}\right]\mathbb{P}\left[\boldsymbol{x}|\theta^{p}\right]}\\
= & \left\langle \nabla_{\theta}\log Q_{\theta,\theta^{p}}\left[\boldsymbol{x}\right]\log\frac{Q_{\theta,\theta^{p}}\left[\boldsymbol{x}\right]}{\mathbb{P}\left[O|\boldsymbol{x},\theta^{p}\right]\mathbb{P}\left[\boldsymbol{x}|\theta^{p}\right]}\right\rangle _{Q_{\theta,\theta^{p}}}\label{eq:parameterlearning}
\end{align}
where in Eq.~\ref{eq:cancelled} we used the fact that $\nabla_{\theta}Q_{\theta,\theta^{p}}=Q_{\theta,\theta^{p}}\nabla_{\theta}\log Q_{\theta,\theta^{p}}$
and that $\int d\boldsymbol{x}Q_{\theta,\theta^{p}}\left[\boldsymbol{x}\right]\nabla_{\theta}\log Q_{\theta,\theta^{p}}\left[\boldsymbol{x}\right]\equiv0$.
\begin{align}
\nabla_{\theta^{p}}\int d\boldsymbol{x}Q_{\theta,\theta^{p}}\left[\boldsymbol{x}\right]\log\frac{Q_{\theta,\theta^{p}}\left[\boldsymbol{x}\right]}{\mathbb{P}\left[O|\boldsymbol{x},\theta^{p}\right]\mathbb{P}\left[\boldsymbol{x}|\theta^{p}\right]}= & \nabla_{\theta^{p}}\int dxQ_{\theta,\theta^{p}}\left[\boldsymbol{x}\right]\log\frac{Q_{\theta,\theta^{p}}\left[\boldsymbol{x}\right]}{\mathbb{P}\left[O|\boldsymbol{x},\theta^{p}\right]\mathbb{P}\left[\boldsymbol{x}|\theta^{p}\right]}\\
= & \int d\boldsymbol{x}\nabla_{\theta^{p}}Q_{\theta,\theta^{p}}\left[\boldsymbol{x}\right]\log\frac{Q_{\theta,\theta^{p}}\left[\boldsymbol{x}\right]}{\mathbb{P}\left[O|\boldsymbol{x},\theta^{p}\right]\mathbb{P}\left[\boldsymbol{x}|\theta^{p}\right]}+\nonumber \\
 & +\int d\boldsymbol{x}Q_{\theta}\left[x\right]\nabla_{\theta^{p}}\log\frac{Q_{\theta,\theta^{p}}\left[\boldsymbol{x}\right]}{\mathbb{P}\left[O|\boldsymbol{x},\theta^{p}\right]\mathbb{P}\left[\boldsymbol{x}|\theta^{p}\right]}\\
= & \int dxQ_{\theta,\theta^{p}}\left[\boldsymbol{x}\right]\nabla_{\theta^{p}}\log Q_{\theta}\left[\boldsymbol{x}\right]\log\frac{Q_{\theta,\theta^{p}}\left[\boldsymbol{x}\right]}{\mathbb{P}\left[O|\boldsymbol{x},\theta^{p}\right]\mathbb{P}\left[\boldsymbol{x}|\theta^{p}\right]}+\nonumber \\
 & -\int dxQ_{\theta,\theta^{p}}\left[\boldsymbol{x}\right]\nabla_{\theta^{p}}\log\mathbb{P}\left[O|\boldsymbol{x},\theta^{p}\right]\mathbb{P}\left[\boldsymbol{x}|\theta^{p}\right]\\
= & \left\langle \nabla_{\theta^{p}}\log Q_{\theta,\theta^{p}}\left[\boldsymbol{x}\right]\log\frac{Q_{\theta,\theta^{p}}\left[\boldsymbol{x}\right]}{\mathbb{P}\left[O|\boldsymbol{x},\theta^{p}\right]\mathbb{P}\left[\boldsymbol{x}|\theta^{p}\right]}\right\rangle _{Q_{\theta,\theta^{p}}}+\nonumber \\
 & -\left\langle \nabla_{\theta^{p}}\log\mathbb{P}\left[O|\boldsymbol{x},\theta^{p}\right]\mathbb{P}\left[\boldsymbol{x}|\theta^{p}\right]\right\rangle _{Q_{\theta,\theta^{p}}}\label{eq:Appendix:HyperparamsKLderivative}
\end{align}

Unsurprisingly, Eq.~\eqref{eq:parameterlearning} is identical to Eq.~\eqref{eq:Appendix:derivativeKLparams}. Eq.~\eqref{eq:Appendix:HyperparamsKLderivative} is similar to Eq.~\eqref{eq:Appendix:derivativeKLparams}, with an extra second term that accounts for the direct dependence of the model on the hyperparameters. Note that this second term can be interpreted as a negated derivative of the energy in an Expectation Maximization (EM) scheme. The learning process
for $\theta^{p}$ is analogous to learning $\theta$: 
the averages in r.h.s of Eq. \eqref{eq:Appendix:HyperparamsKLderivative}
are calculated by sampling and then the hyperparameters are updated using the Sign Descender rule. From a computational point of view, it is convenient to update both the parameters $\theta$ and the hyperparameters $\theta^p$ at each iteration.

\section{Other Inferential Methods \label{sec:inferential_methods}}
This section provides a brief description of the other inferential techniques whose performances are compared with those of \Causalityname{} in Section \ref{sec:EpidemicRiskAss} in the main text.
\paragraph{Sib --}
This method is based on a Belief Propagation approach to epidemic spreading processes, that allows one to compute, in an efficient and distributed way, the marginals over a posterior distribution (Eq. \eqref{eq:post}) for compartmental epidemic models with non-recurrent dynamics (eventually non-Markovian). It shows very good performances when epidemic models take place on random contact networks, while it may suffer from the presence of loops in the graph. A detailed explanation of this method can be found in  \cite{baker_epidemic_2021}.

\paragraph{Mean Field (MF) --}
This method is based on a heuristic way to deal with observations from clinical tests and on a Mean Field approximation of the prior distribution, which is considered to be factorized over nodes at each time. An advantage of this method relies on its simplicity and small computational cost; however, it typically shows poorer performances with respect to the other methods. We refer to \cite{baker_epidemic_2021} for additional details about the MF approximation. The heuristic scheme is instead discussed in the next paragraph.
\paragraph{Heuristic (heu) --}
The MF method developed in \cite{baker_epidemic_2021} and described above relies on two approximations. The first is a heuristic way to deal with observations: it consists in assuming that if an individual has tested positive at time $t$, then it became positive at time $t-\tau$, with $\tau$ properly tuned (to further details we refer to \cite{baker_epidemic_2021}). The second is an MF ansatz for the prior distribution. It is natural to ask whether such a heuristic is a good approximation for the observation constraints, regardless of the MF factorization ansatz. We, therefore, replace the MF estimation of marginal probabilities by sampling trajectories forward in time. 
As shown in Section \ref{sec:EpidemicRiskAss}, this method shows slightly better performances w.r.t. MF. The heuristic, therefore, is a good guess for epidemic risk assessment.
As a final remark, the \Causalityname{} can actually be considered as a further extension of the MF approximation \cite{baker_epidemic_2021} in the following sense:
\begin{enumerate}
\item it gives a full justification to the heuristic using a variational
principle;
\item it extends \cite{baker_epidemic_2021} by allowing the patient
zero inference;
\item it allows one to infer the hyperparameters of the generative
model and to perform epidemic reconstruction in the case of non-Markovian dynamics.
\end{enumerate}

\paragraph{Monte Carlo}
The results obtained for the SI model in Figs. \ref{fig:The-zero-patient}-\ref{fig:Tubingen} in the main text are also compared to a standard Markov-Chain Monte-Carlo (MCMC) sampling for the posterior distribution. Since epidemic trajectories can be fully described in terms of the infection time vector $\boldsymbol{t}=\left(t_1, \ldots, t_N\right)$, the Markov chain defines dynamics on these continuous variables that eventually converge to a stationary distribution (i.e. the posterior). At each step of the MCMC, a node $i$ is randomly selected and a new value of its infection time - denoted with $t'_i$ - is proposed, by drawing it from a suitable kernel $\mathcal{K} \left(t'_i \mid t_i \right) $.
We adopted a Gaussian kernel centered at $t_i$, with fixed standard deviation $\delta$. The proposed value $t'_i$ sampled from $\mathcal{K}$ is then clamped within the interval $\left[0,T\right]$: this procedure allows to sample with non-zero probability the two extreme values, associated to node $i$ being a patient-zero (when $t_i=0$), or node $i$ never being infected (in which case its infection time is formally set to infinity, as previously discussed). This procedure is equivalent to use as proposal kernel a 2-sided rectified Gaussian distribution in the window $\left[0,T\right]$.
The acceptance probability of such a proposal is computed as
\begin{equation}
    p_{\text{acc}} = \min \left(1, \frac{\mathbb{P}\left[\boldsymbol{t}'_i|\mathcal{O}\right] }{\mathbb{P}\left[\boldsymbol{t}|\mathcal{O}\right]} \frac{\tilde{\mathcal{K}}\left(t_i \mid t'_i,\delta \right)}{\tilde{\mathcal{K}}\left(t'_i \mid t_i,\delta \right)} \right)
\end{equation}
where $\boldsymbol{t}'_i$ represents an epidemic trajectory where only $t_i$ is modified to the corresponding proposed value, namely $\boldsymbol{t}'_i = \left( t_1, \ldots, t'_i, \ldots, t_N \right)$, and $\tilde{\mathcal{K}}$ is the rectified kernel given by:
\begin{equation}
\tilde{\mathcal{K}} \left(x\mid t_{i},\delta\right)=\Phi\left(0;t_i, \delta\right)\delta\left(x\right)+\mathbb{I}\left[x \in \left(0,T\right)\right]K\left(x;t_{i},\delta \right) +\left[1-\Phi\left(T; t_i, \delta\right)\right]\delta\left(x-T\right)
\end{equation}

with $K \left(x;\mu, \sigma \right) $ being the probability density of a Gaussian random variable and $\Phi \left(x; \mu, \sigma \right) $ its cumulative function.
The initial condition for the Markov Chain is sampled from the prior distribution of the SI model. In order to have a fair comparison, the number of samples collected during the MCMC is equal to the one used by \Causalityacronym{}. To diminish the effect of initial equilibration time an initial number of steps is typically required to let the MC forget the initial condition and sample efficiently the posterior distribution; notice that a ``step'' consists in proposing a move for each node in a random permutation. 

\paragraph{Soft-Margin --}
The Soft-Margin estimator is described in \cite{softmarg}. We adapted this method by sampling from the prior probability distribution $\mathbb{P}[\boldsymbol{x}]$ and weighting each sample with the observation likelihood $\mathbb{P}[\mathcal{O}|\boldsymbol{x}]$, in which we introduced a small artificial noise in the form of a false rate, which softened the constraints, improving the performances of the method. The technique is asymptotically exact. However, when the population size grows, the probability to sample a trajectory $\boldsymbol{x}$ which satisfies the observation constraints $\mathbb{P}[\mathcal{O}|\boldsymbol{x}]$ dramatically decreases. Therefore, we expect the method to be quite slow for large population sizes in order to well-approximate the posterior distribution. 

\section{Models of contact networks used in the numerical simulations \label{sec:contact_networks}}
\paragraph{Proximity model --}
The proximity contact graph is obtained by distributing $N$ individuals uniformly in a square of side $\sqrt{N}$ and generating the contacts as follows. A contact can be established between two individuals $i$ and $j$ with a probability $e^{ -d_{ij} / \ell  }$, where $d_{ij}$ is the Euclidean distance between the points $i$ and $j$ are located and $\ell$ is a length scale that can be tuned to change the density of the contact graph. 

\paragraph{OpenABM-Covid19 --}
A synthetic contact network has been generated through OpenABM, a platform used to set up realistic epidemic instances on dynamic contact networks where a set of intervention measures have been applied and studied in Ref. ~\cite{hinch2021openabm}. The underline contact pattern is a superposition of two static graphs, one a complete graph representing interaction within households and a second small-word network mirroring occupation relationships. A random time-varying network is instead used and rebuilt on a daily basis to model contacts in public transportation, transient social gatherings, etc. The number of interactions through the random network is extracted from a negative binomial distribution to allow for rare super-spreading events. Memberships to both fixed and dynamic graphs are determined by the age of individuals, i.e. children live with adults, elderly people have fewer interactions than other age groups, etc. \\
In this work, we generated an instance of $N=1000$
individuals using default parameters.

\paragraph{Spatio-temporal model from geolocation data --}

Dynamic contact network instances with realistic patterns of time-dependent contacts can be generated using the spatio-temporal model in Ref. ~\cite{lorch2022quantifying}. 
According to this model, individuals are assigned to households that are localized in an urban area according to the actual population density. According to available geo-location data, other venues (schools and research institutes, social places, bus stops, workplaces, and supermarkets) are similarly displaced in the map. 
In a mobility simulation, individuals can visit a  number of locations with a probability that decreases as the household-target distance increases. The duration of contacts between individuals concurrently visiting the same venue is assumed to be known and gathered by contact tracing smartphone applications. Some interesting and realistic features naturally arise from this contact dynamics, such as the presence of super-spreaders, e.g. the number of infections caused by infectious individuals is over-dispersed.
In the present work, only a small example of such kind of dynamically generated contact network was analyzed. The contact-network instances used to obtain results in Fig.~\ref{fig:Tubingen} in the main text represent the interaction graph among a small community of  $N = 904$ individuals moving on a coarse-grained version of the city of T\"{u}bingen, in Germany. Due to the rescaling of the population, also the number of venues in the urban area was rescaled by a factor of 20. 

\section{Proof that conditioning a Markov process with local time observations leads to a Markov process}
Let us suppose to have a Markov process described by the following probability distribution:
\begin{equation}
\mathbb{P}[\mathbf{x}]=\mathbb{P}[x({0})]\prod_{t=1}^{T-1}\mathbb{P}[x({t})|x({t-1})]
\end{equation}
Now we constrain the dynamics to a set $\mathcal{O}=((O_1,t_1)\dots,(O_M,t_M))$ of local observations in time, i.e. with probability law $\mathbb{P}[\mathcal O|x(0),\dots,x(T)]=\prod_{t}\prod_{\mu:t_{\mu}=t} \mathbb{P}[O_\mu|x(t)]=:\prod_{t} \mathbb{P}[O_t|x(t)]$.  We will show that the posterior distribution is still a Markov process, i.e.
\begin{align}
    \mathbb{P}[x(0),\dots,x(T)|\mathcal O] = \mathbb{P}[x(0)|\mathcal O]\prod_{t=1}^T \mathbb{P}[x({t})|x({t-1}),\mathcal O]
\end{align}

The posterior probability
can be rewritten using Bayes' Theorem as:
\begin{equation}
\mathbb{P}[\mathbf{x}|\mathcal{O}]\propto\mathbb{P}[\mathbf{x}]\mathbb{P}[\mathcal{O}|\mathbf{x}]=\mathbb{P}[x({0})]\left(\prod_{t=0}^{T-1}\mathbb{P}[x({t+1})|x({t})]\right)\left(\prod_{t=0}^{T}\mathbb{P}[\mathcal{O}_{t}|x({t})]\right),
\end{equation}
where we exploited that the observations are local in time and therefore the conditional probability factorizes.
From the above equation we can try to calculate the posterior probability at a given time $s$ conditioned to the whole past:
\begin{align}
\mathbb{P}[x({s})|x({0}),x({1}),\dots,x({s-1}),\mathcal{O}] & =\frac{\mathbb{P}[x({s}),x({0}),x({1}),\dots,x({s-1})|\mathcal{O}]}{\mathbb{P}[x({0}),x({1}),\dots,x({s-1})|\mathcal{O}]} \nonumber \\
 & \propto{\mathbb{P}[x({s}),x({0}),x({1}),\dots,x({s-1})|\mathcal{O}]}  \nonumber \\
& =\sum_{x({s+1}),\dots,x({T})}\mathbb{P}[x(0),\dots,x(T)|\mathcal{O}] \nonumber \\
 & =\mathbb{P}[x({0})]\left(\prod_{r=0}^{s-2}\mathbb{P}[x({r+1})|x({r})]\right)\left(\prod_{r=0}^{s-1}\mathbb{P}[\mathcal{O}_{r}|x_{r}]\right) f\left( x({s}),x({s-1}), \mathcal{O} \right) \nonumber \\
 & \propto f(x({s}),x({s-1}),\mathcal{O})
 . \label{eq:posterior_conditionedonpast}
\end{align}
where the symbol $\propto$ is intended with respect to $x_s$ so all terms that do not depend on $x_{s}$ can be removed and we defined:
\begin{equation}
f(x({s}),x({s-1}),\mathcal{O}):=\mathbb{P}[x({s})|x({s-1}),\mathcal{O}]\sum_{x({s+1}),\dots,x({T})}\left(\prod_{r=s}^{T-1}\mathbb{P}[x({r+1})|x({r})]\right)\left(\prod_{r=s}^{T}\mathbb{P}[\mathcal{O}_{r}|x({r})]\right),
\end{equation}
Now:
\begin{align}
\mathbb{P}[x({s})|x({s-1}),\mathcal{O}]
&=\sum_{x({s-2}),\dots,x(0)} \mathbb{P}[x({s})|x({s-1}),\dots,x(0),\mathcal{O}] \mathbb{P}[x({s-1}),\dots,x(0)|\mathcal{O})\\
&\propto f(x({s}),x({s-1}),\mathcal{O})\sum_{x({s-2}),\dots,x(0)}\mathbb{P}[x({s-1}),\dots,x(0)|\mathcal{O}]\\
&\propto f(x({s}),x({s-1}),\mathcal{O})\\
&\propto \mathbb{P}[x({s})|x({s-1}),\dots,x(0),\mathcal{O}]
\end{align}
Where again, any term that is constant with respect to $x_s$ was removed. As $\mathbb{P}[x({s})|x({s-1}),\mathcal{O}] \propto \mathbb{P}[x({s})|x({s-1}),\dots,x(0),\mathcal{O}]$ and both are normalized with respect to $x_s$, they must be identical. Therefore, using equation \eqref{eq:causal_post} we conclude the proof.
We can conclude that, at fixed observations, the posterior distribution at a given time $s$ can be represented as a Markov process, i.e. depending only on the previous time $s-1$. However, the transition probabilities between these two times will depend on the observations at future times, i.e. $t\geq s$ and a closed formula would require to recast $f\left(x(s),x(s-1) \mid \mathcal{O}\right) =\tilde{\mathbb{P}} [x(s) | x(s-1), \mathcal{O}] $,
which involves an exponential number of terms to be computed (remember that $x$ is the full state of all the degrees of freedom). In this perspective, the CVA further assumes that even these transition probabilities factorize, i.e. there is a spatial conditional independence over the posterior, a necessary ingredient for sampling efficiency.

\end{document}